\documentclass[a4paper,11pt]{article}
\pdfoutput=1 

\usepackage{jheppub} 

\usepackage[T1]{fontenc} 

\usepackage{amsfonts,amsmath,amssymb}
\usepackage{graphicx} 
\usepackage{booktabs}
\usepackage{hyperref}
\usepackage{mathtools}
\usepackage{float}
\usepackage[nottoc, notlot, notlof]{tocbibind}
\usepackage{physics}
\usepackage{bigints}
\usepackage[makeroom]{cancel}
\usepackage{braket}
\usepackage[T1]{fontenc}
\usepackage{color}
\usepackage{graphicx}
\usepackage{float}
\usepackage{chngpage}
\usepackage{wrapfig}

\allowdisplaybreaks

\title{\boldmath Resummation methods for Master Integrals}

\author[a,b]{Dhimiter D. Canko}
\author[a,c]{and Nikolaos Syrrakos}
\affiliation[a]{Institute of Nuclear and Particle Physics, NCSR "Demokritos",\\ Agia Paraskevi 15310, Greece}
\affiliation[b]{Department of Physics, University of Athens,\\Zographou 15784, Greece}
\affiliation[c]{Physics Division, National Technical University of Athens, \\Athens 15780, Greece}
\emailAdd{jimcanko@phys.uoa.gr}
\emailAdd{syrrakos@inp.demokritos.gr}

\abstract{We present in detail two resummation methods emerging from the application of the Simplified Differential Equations approach to a canonical basis of master integrals. The first one is a method which allows for an easy determination of the boundary conditions, since it finds relations between the boundaries of the basis elements and the second one indicates how using the $x \rightarrow 1$ limit to the solutions of a canonical basis, one can obtain the solutions to a canonical basis for the same problem with one mass less. Both methods utilise the residue matrices for the letters $\{0,1\}$ of the canonical differential equation. As proof of concept, we apply these methods to a canonical basis for the three-loop ladder-box with one external mass off-shell, obtaining subsequently a canonical basis for the massless three-loop ladder-box as well as its solution.}


\begin{document} 
\maketitle
\flushbottom

\section{Introduction}
\label{sec:intro}

The forthcoming update of the LHC to its High-Luminosity phase will usher in an era of increased experimental precision. This poses a challenge from a theoretical standpoint, as predictions of scattering processes of matching precision to the experimental one will be crucial, in order to find subtle effects of New Physics hidden within the data.

A key ingredient for the calculation of the outcome of a scattering process in Perturbative Quantum Field Theory is the scattering amplitude. Within these amplitudes, the main actors are the well-known Feynman Diagrams, which can be associated to their corresponding Feynman Integrals (FI) in a systematic way. A lot of progress has happened in the last decade in the methods that are employed for the calculation of FI, both numerically and analytically. However, with each integral that one calculates and with each problem that gets solved, a new one arises which challenges the current methods. 

A method, working within dimensional regularization and applied for \textit{families} of FI\footnote{Each Feynman Integral can be associated with a general class of integrals, called a \textit{family}.}, which has provided many important analytical results is the method of Differential Equations (DE) \cite{de1,de2,de3,de4}. This method takes advantage of the fact that FI of a family are not independent due to Integration-By-Parts relations (IBP) \cite{IBP1} but can be written in terms of a linear combination of a minimal finite (not unique) basis of integrals, known as Master Integrals (MI). Many packages have been developed \cite{kira,fire,litered} based on Laporta's algorithm \cite{IBP2} that implement the method of reduction to MI via IBP relations. Nevertheless, the procedure of reduction still remains a difficult problem from a computational point of view for families with many loops and/or many kinematic invariants, leading to the search for new methods \cite{IBP3,IBP4,IBP5,IBP6,IBP7,IBP8,IBP9,IBP10,IBP11}. The method of DE lived through a revolution in recent years with the introduction of the concept of \textit{universally transcendental and pure} functions and the so-called \textit{canonical form} of the DE \cite{Henn1}. Several packages \cite{Henn2,CB1,CB2,CB3,CB4,CB5,CB6,CB7,CB8,CB9} have been developed for the determination of a canonical basis of MI, which satisfy a canonical DE, however there is an ongoing effort to optimise the currently used algorithms and develop new ones.  

A variant of the DE method, known by now as the Simplified Differential Equations approach (SDE) \cite{SDE1,SDE2,SDE3}, has also provided several important results. A distinct advantage of the SDE approach is the fact that regardless of the number of scales of the problem at hand, the resulting DE have only one variable. Combining the SDE approach with the ideas of \cite{Henn1}, yields a canonical DE in one variable with purely numerical residue matrices. Integrating iteratively for each order of the dimensional regulator yields a solution written in a very compact form, where the main inputs are the residue matrices corresponding to each letter and some boundary terms. 

In this paper we give a detailed account on the resummation method used to determine the boundary terms \cite{5box}. This method is based on the SDE approach, by exploiting the asymptotic behaviour of the basis elements at $x=0$, so that the result can be readily expressed in terms of Goncharov Polylogarithms (GPLs) \cite{GPL}. Another key feature of the SDE approach is that by carefully taking the $x\to 1$ limit \cite{SDE3, Papadopoulos:2019iam} of the solution for a problem with $n$ external masses, one can obtain the solution to the problem with $n-1$ external masses. In the second half of this paper, we present a method for the determination of a canonical basis of MI and its solution, itself expressed in terms of GPLs, for a problem with massless external legs, once a canonical basis and its solution for a problem with one massive external leg is known. 

Both of these methods, mentioned above, utilise the residue matrices of the letter $\{0\}$ for the former and $\{1\}$ for the latter. By applying the process of Jordan Decomposition \cite{5box, Papadopoulos:2019iam} to these purely numerical matrices, we may define their corresponding resummation matrices. These resummation matrices are the key players in the methods presented in this paper.  As a new non-trivial application of these methods, we choose the three loop ladder-box family, both with one \cite{mastrolia} and zero \cite{Henn:2013tua} external masses, thus establishing the applicability of the SDE approach for the computation of three loop MI. Both families of FI have already been calculated, however for the one-mass family we report a small correction in the number of MI to the previously published result and for the massless one we provide an original canonical basis of MI. 

The outline of this paper is as follows. In section \ref{sec:setup} we set up the basic formalism for the problem that we will discuss. In section \ref{sec:boundaries} we give a detailed account of the techniques employed to find the necessary boundary terms for the solution of the three loop ladder-box family with one external mass with an emphasis on the resummation method at $x=0$. In section \ref{sec:massless} we analyse how to take the $x\to 1$ limit of our solution for the massive family and we show how this can yield a canonical basis for the massless family. Finally in section \ref{sec:conlusions} we summarise our results and give a perspective look to the future use of these methods.

All diagrams have been created using \texttt{jaxodraw} \cite{jaxodraw} and throughout our calculations we have made extensive use of the programs \texttt{FIRE6} \cite{fire} and \texttt{KIRA} \cite{kira} for the IBP reduction and of the packages \texttt{PolyLogTools} \cite{Polylogtools} and \texttt{HyperInt} \cite{hyper} for the manipulation of the resulting GPLs.

\section{Three-loop ladder-box with one external mass}
\label{sec:setup}

The FI family of the three-loop ladder-box with one external massive leg, assuming all the external particles as incoming, is described by the Feynman graph pictured in \textbf{Figure 1}, and was first studied in \cite{mastrolia}. We adopt the notation and the canonical basis presented within \cite{mastrolia}. The latter can be found, in the SDE notation, in \textbf{Table 1} and \textbf{Table 2} at the end of the article.

\begin{figure} [h!]
\centering
\includegraphics[width=4 in]{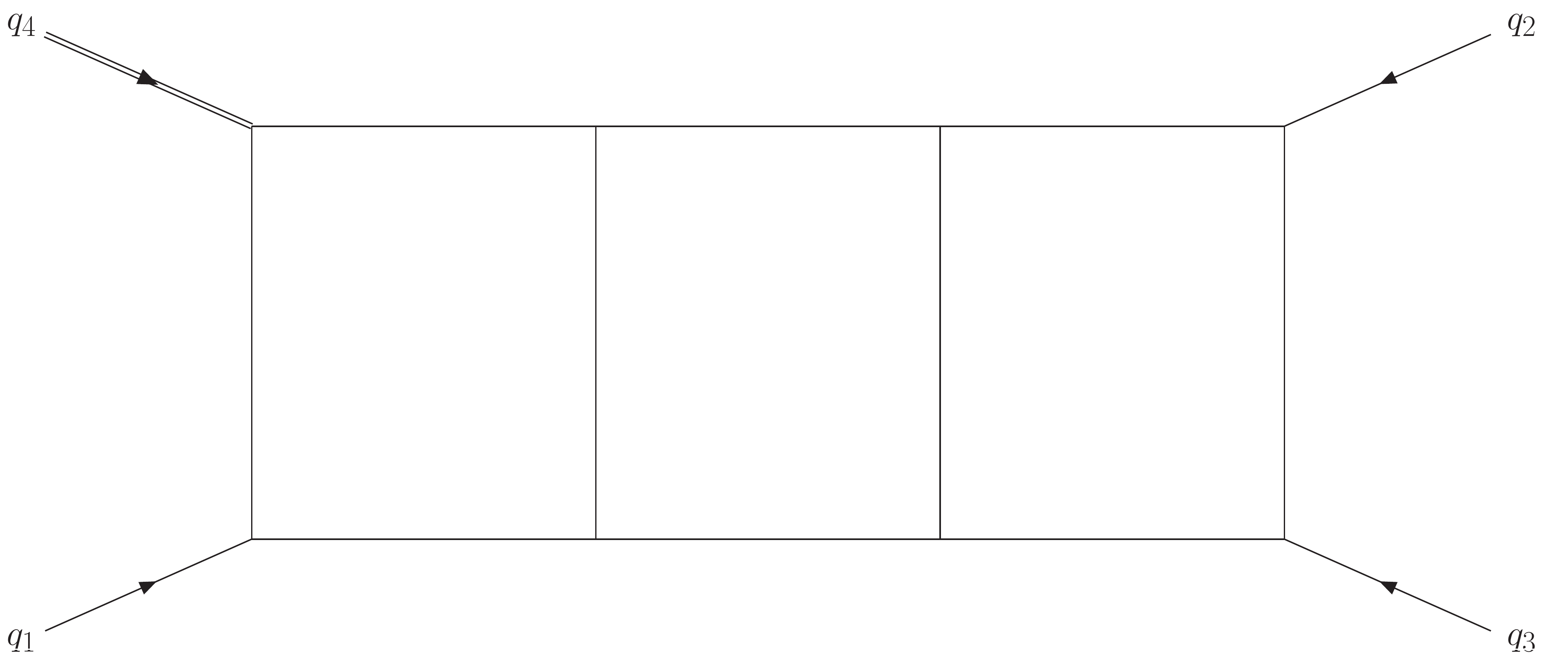}
\caption{The three-loop ladder-box family with one external massive leg}
\end{figure}

The class of FI describing this family can be expressed as follows
\begin{equation}
\label{family}
G_{a_1,\dots,a_{15}}\left( \{ q_j \}, \varepsilon \right)=\int \left( \prod_{r=1}^3 \frac{d^d l_r}{i \pi^{d/2}} \right) \frac{e^{3 \varepsilon \gamma_E}}{D_{1}^{a_1} \dots D_{15}^{a_{15}}} \, \, \, \, \, \, \, \text{with} \, \, \, \, \, \, \, d=4-2\varepsilon \, ,
\end{equation}
where the chosen parametrization for the propagators is\footnote{we use the abbreviation $q_{12}= q_1 + q_2$ and $q_{123}=q_1+q_2+q_3$}
\begin{equation}
\label{propagators}
\begin{split}
&D_1=l_1^2 \, , \, \, \, \, \, \, \, D_2=l_2^2 \, , \, \, \, \, \, \, \, D_3=l_3^2 \, , \, \, \, \, \, \, \, D_4=(l_1-l_2)^2 \, , \, \, \, \, \, \, \, D_5=(l_2-l_3)^2 \, ,\\
&D_6=(l_3+q_2)^2 \, , \, \, \, \, \, \, \, D_7=(l_1+q_{23})^2 \, , \, \, \, \, \, \, \, D_8=(l_2+q_{23})^2 \, , \, \, \, \, \, \, \, D_9=(l_3+q_{23})^2 \, ,\\
& D_{10}=(l_1+q_{123})^2 \, , \, \, \, \, \, \, \, D_{11}=(l_1+q_2)^2 \, , \, \, \, \, \, \, \, D_{12}=(l_2+q_2)^2 \, , \, \, \, \, \, \, \, D_{13}=(l_2+q_{123})^2 \, ,\\
&D_{14}=(l_3+q_{123})^2 \, , \, \, \, \, \, \, \, \text{and} \, \, \, \, \, \, \, D_{15}=(l_1-l_3)^2 \, ,
\end{split}
\end{equation} 
and the external momenta obey the following kinematics
\begin{equation}
\label{invariants}
q_1^2=q_2^2=q_3^2=0 \, , \, \, \, \, \, \, \, q_4^2=m^2 \, , \, \, \, \, \, \, \, q_2 \cdot q_3=s/2 \, , \, \, \, \, \, \, \, q_1 \cdot q_3=t/2 \, , \, \, \, \, \, \, \, q_1 \cdot q_2=(m^2 - s - t)/2 \, .
\end{equation} 
In (\ref{family}) $D_{11}, \dots, D_{15}$
are not uniquely defined propagators coming from irreducible scalar products and thus for their corresponding integers we have $\{a_{11},a_{12},a_{13},a_{14},a_{15}\}\leq 0$.

Within the SDE approach, we insert a dimensionless parameter, $x$, which simplifies drastically the derivation of the required differential equations. In this case, we choose the following parametrization
\begin{equation*}
\label{SDE}
q_1 \rightarrow x p_1 \, , \, \, \, \, \, \, \, q_2 \rightarrow p_3 \, , \, \, \, \, \, \, \, q_3 \rightarrow -p_{123} \, , \, \, \, \, \, \, \, q_4 \rightarrow p_{12} -x p_1 \, \, \, \, \, \, \, \text{with} \, \, \, \, \, \, \, p_1^2=p_2^2=p_3^2=p_4^2=0 \, .
\end{equation*}

\begin{figure} [h!]
\centering
\includegraphics[width=4 in]{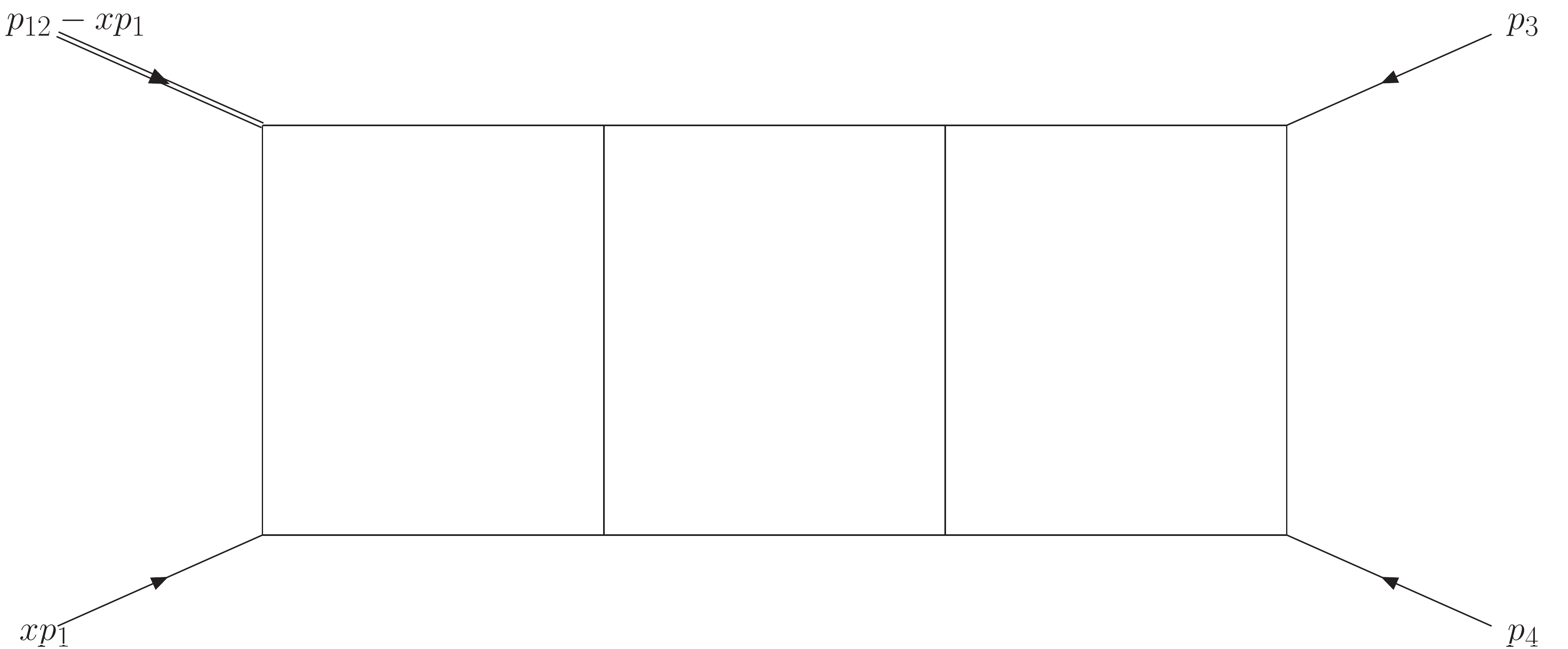}
\caption{The three-loop ladder-box family with one external massive leg in the SDE parametrization.}
\end{figure}

This parametrization allows us to express the Mandelstam variables and the external mass defined in (\ref{invariants}) in terms of the parameter $x$ and the new Mandelstam variables of the null momenta $p_j$
\begin{equation}
\label{SDEinva}
s=s_{12} \, , \, \, \, \, \, \, \, t=x s_{23} \, , \, \, \, \, \, \, \, m^2=(1-x) s_{12} \, ,
\end{equation} 
where $s_{12}=(p_1 + p_2)^2$ and $s_{23}=(p_2 + p_3)^2$. Furthermore, the propagators defined in (\ref{propagators}) after making the transformations $(l_1 \rightarrow k_1-q_{23}, \, l_2 \rightarrow -k_2-q_{23}, \, l_3\rightarrow k_3-q_{23})$ to the loop momenta and applying the SDE approach take the following form in the new notation
\begin{equation}
\label{SDEprop}
\begin{split}
&D_1=(k_1+p_{12})^2 \, , \, \, \, \, \, \, \, D_2=(k_2-p_{12})^2 \, , \, \, \, \, \, \, \, D_3=(k_3+p_{12})^2 \, , \, \, \, \, \, \, \, D_4=(k_1+k_2)^2 \, ,\\
& D_5=(k_2 + k_3)^2 \, , \, \, \, \, \, \, \, D_6=(k_3 + p_{123})^2 \, , \, \, \, \, \, \, \, D_7=k_1^2 \, , \, \, \, \, \, \, \, D_8=k_2^2 \, , \, \, \, \, \, \, \, D_9=k_3^2 \, ,\\
& D_{10}=(k_1 + x p_1)^2 \, , \, \, \, \, \, \, \, D_{11}=(k_1 + p_{123})^2 \, , \, \, \, \, \, \, \, D_{12}=(k_2 - p_{123})^2 \, ,\\
&D_{13}=(k_2 - x p_1)^2 \, , \, \, \, \, \, \, \, D_{14}=(k_3 +x p_1)^2 \, , \, \, \, \, \, \, \, \text{and} \, \, \, \, \, \, \, D_{15}=(k_1 - k_3)^2 \, .
\end{split}
\end{equation} 

Performing the reduction to MI, we found for this family a set of 83 MI in contrast with \cite{mastrolia}, where a set of 85 MI was presented. The two extra MI contained in the set of 85 MI were found to be equal from IBP relations with two other integrals of the same set, namely $\mathcal{T}_7 = \mathcal{T}_8$ and $\mathcal{T}_{45} = \mathcal{T}_{46}$ of \cite{mastrolia}. These relations can also be verified by checking the solutions for the corresponding basis elements, as presented in \cite{mastrolia}.

Regarding the IBP reduction that we performed, we should note here that to speed up the reduction and obtain the DE with respect to $x$ we did the reduction keeping only $x$ as a variable and choosing different numerical values (high prime numbers) for $s_{12}$ and $s_{23}$. This procedure was made possible by the fact that we already knew one canonical basis and the letters of the DE with respect to $x$ for this family, which are the same with the letters from the family of the double-box with one external massive leg.

Having a canonical basis for the studied family we obtained a DE with respect to $x$ which is of canonical form
\begin{equation}
\label{de}
\partial_{x} \textbf{g}=\varepsilon \left( \sum_{i=1}^4 \frac{\textbf{M}_i}{x-l_i} \right) \textbf{g}
\end{equation} 
with $\textbf{M}_i$ being purely numerical matrices and $l_i=\{0,\,1,\, s_{12}/(s_{12}+s_{23}),-s_{12}/s_{23}\}$ the letters of the alphabet. We solve the DE for the basis elements up to weight six in the Euclidean region of the invariants, where FI are free of branch cuts. In our case, we find that the Euclidean region in the SDE notation is 
\begin{equation}
\label{euclidean}
0<x<1\, , \, \, \, \, \, \, \, s_{12}<0 \, , \, \, \, \, \, \, \, s_{12}<s_{23}<0 \, .
\end{equation} 
The solution of \eqref{de} can be written in the compact form
\begin{equation}
\label{solution}
\begin{split}
\textbf{g}&=\varepsilon^0 \textbf{b}_0^{(0)}+\varepsilon \left(\sum {\cal G}_i \textbf{M}_i \textbf{b}_0^{(0)}+\textbf{b}_0^{(1)}\right)+\varepsilon^2 \left(\sum {\cal G}_{ij} \textbf{M}_i\textbf{M}_j\textbf{b}_0^{(0)}+\sum {\cal G}_i \textbf{M}_i \textbf{b}_0^{(1)}+\textbf{b}_0^{(2)} \right)+ \dots \\  
&+ \varepsilon^6 \left(\textbf{b}_0^{(6)}+ \sum {\cal G}_{ijklmn} \textbf{M}_i \textbf{M}_j \textbf{M}_k \textbf{M}_l \textbf{M}_m \textbf{M}_n \textbf{b}_0^{(0)} + \sum {\cal G}_{ijklm} \textbf{M}_i \textbf{M}_j \textbf{M}_k \textbf{M}_l \textbf{M}_m \textbf{b}_0^{(1)} \right. \\
&+\left. \sum {\cal G}_{ijkl} \textbf{M}_i \textbf{M}_j \textbf{M}_k \textbf{M}_l \textbf{b}_0^{(2)} +\sum {\cal G}_{ijk} \textbf{M}_i\textbf{M}_j\text{M}_k \textbf{b}_0^{(3)}+\sum {\cal G}_{ij} \textbf{M}_i\textbf{M}_j\textbf{b}_0^{(4)}+\sum {\cal G}_i \textbf{M}_i \textbf{b}_0^{(5)} \right) \, ,
\end{split}
\end{equation}
where the matrices $\textbf{b}_0^{(i)}$ are the boundary terms, which we will explain in the next section how they were found, and ${\cal G}_i, ... ,{\cal G}_{ijklmn}$ are GPLs of weight $1, \dots,6$, respectively, with argument $x$ and letters from the set $l_i$. Our results for the basis elements can be found in the ancillary files together with the matrices of the DE. These were crossed-checked numerically with the results from \cite{mastrolia}, and perfect agreement was found in all cases.

\section{Boundary Conditions}
\label{sec:boundaries}

In this section we describe the methods we used to compute the boundary conditions of the basis elements at $x=0$,  with emphasis on the resummation method \cite{5box} which allows us to obtain relations between the boundary conditions of different basis elements. This result is very important as it reduces a lot the number of unknown boundaries and moreover gives results for boundaries which otherwise would have been difficult to compute with other methods.

\subsection{Known and zero boundaries}

Some integrals of this family appear in the family of the massless three-loop ladder-box. Because of that, we can obtain results and therefore boundary conditions, for these integrals from the ones of the massless problem \cite{Henn:2013tua}. Hence we right away obtain boundaries for the following basis elements (from now on we will use the notation $gb_i$ for the boundaries of the basis elements)
\begin{equation*}
\{ gb_1, \, gb_2, \, gb_3, \, gb_4, \, gb_5, \, gb_6, \, gb_7, \, gb_{17}, \, gb_{18}, \, gb_{19}, \, gb_{44} \} \, .
\end{equation*}

Beyond the knowledge of the above boundaries, we can apply the method of expansion-by-regions \cite{asy1,asy2,asy3,asy4,asy5,asy6} and utilise the information of the asymptotic limit of the MI at $x \rightarrow 0$ and of the form with which they appear in the basis elements (see \textbf{Table 1} and \textbf{Table 2}) to set some boundaries equal to zero\footnote{This information can be also obtained for some basis elements from the method described in the next subsection.}. More specifically, if the basis element has as an overall prefactor of $x$ in such a power such as its leading regions contributing to its asymptotic limit $x \rightarrow 0$ are of the form $x^{\alpha + \beta \varepsilon}$ with $\alpha>0$, then its boundary term should vanish. Using this observation we found that the following basis elements have zero boundary conditions
\begin{equation*}
\begin{split}
\{ &gb_{10}, \, gb_{11}, \, gb_{14}, \, gb_{15}, \, gb_{21}, \, gb_{22}, \, gb_{23}, \, gb_{24}, \, gb_{25}, \, gb_{26}, \, gb_{28}, \, gb_{31}, \, gb_{37}, \, gb_{38}, \, gb_{45}, \, gb_{46}, \\ \, &gb_{47}, gb_{48}, gb_{50}, \, gb_{53}, \, gb_{55}, \, gb_{58}, \, gb_{59}, \, gb_{63}, \, gb_{64}, \, gb_{66}, \, gb_{68}, \, gb_{70}, \, gb_{80}, \, gb_{82}, \, gb_{83} \} \, .
\end{split}
\end{equation*}
Thus from the $83$ boundary conditions that we need in order to solve the problem at hand, we know from the beginning $42$ of them. For the rest of them, we use the methods described in the next subsections.

\subsection{Relations between boundaries} \label{3.2}

The resummation method utilizes the residue matrix of the canonical differential equation at $x=0$, $\textbf{M}_0$, plus some input from the method of expansion-by-regions \cite{asy1,asy2,asy3,asy4,asy5,asy6} which has been implemented in the \texttt{Mathematica} package \texttt{asy}. The last version of the latter comes together with \texttt{FIESTA4} \cite{fiesta}. Also in order to obtain some actual results, it is necessary to already have a set of known boundary conditions as it happens in our case\footnote{In fact this is always the case as in most families some integrals are already known in closed form or known from families with fewer scales.}.  To make use of this method, we have to study the problem at the integral basis level. To that end, we defined the resummation matrix at $x=0$, $\textbf{R}_0$, through the Jordan-decomposition of $\textbf{M}_0$ \cite{5box}
\begin{equation}
\label{R0}
\begin{split}
\textbf{M}_0 &=\textbf{S}_0 \textbf{D}_0 \textbf{S}^{-1}_0 \, , \\
\textbf{R}_0 &=\textbf{S}_0 e^{\varepsilon \textbf{D}_0 \log(x)} \textbf{S}_0^{-1} \, .
\end{split}
\end{equation} 
This matrix, $\textbf{R}_0$, when acting on the array of the basis elements correctly resumms the logarithms of $x$, in the sense that
\begin{equation}
\label{Greg0}
\textbf{g}=\textbf{R}_0 \textbf{g}_{\text{reg0}} \, ,
\end{equation} 
where $\textbf{g}_{\text{reg0}}$ is the regular part of the basis element at $x=0$. We have subsequently checked that this holds true for our solution.

The asymptotic boundaries at $x=0$ we are seeking for are defined through $\textbf{g}_{\text{reg0}}$ via the relation
\begin{equation}
\label{Gtrunc}
\textbf{g}_{\text{bound}}=\left. \textbf{g}_{\text{reg0}} \right|_{x=0} \, .
\end{equation} 
Multiplying the resummation matrix from the right with $\textbf{g}_{\text{bound}}$ and from the left with the inverse matrix of the transformation that takes us from the MI to the U.T. basis elements, $\textbf{T}^{-1}$, we obtain the asymptotic limit at $x \rightarrow 0$ of the MI \cite{5box}
\begin{equation}
\label{boundaries}
\textbf{F}_{x \rightarrow 0}=\textbf{T}^{-1} \textbf{R}_0 \textbf{g}_{\text{bound}} \, ,
\end{equation} 
which should be equal to the asymptotic limit found for the MI by \texttt{asy}. Through (\ref{boundaries}) we obtain relations between different boundaries of the basis elements by comparing the LHS found by \texttt{asy} with the RHS found by the resummation matrix method.

Let's see now some examples of how this method works. As a first example, we study the master integral $F_{71}$. The expansion-by-regions method yields the following regions that contribute to its result for the limit $x \rightarrow 0: x^{-1-3 \varepsilon}, x^0$ and $x^{-3 \varepsilon}$ . Going back to the resummation matrix for this integral, we notice that it has produced two additional regions: $x^{-1-2\varepsilon}$ and $x^{-1}$. We proceed by setting these regions to zero since they are not predicted by the expansion-by-regions. From the second one, we obtain a relation which connects the boundary condition of $g_{71}$ with the boundary condition of lower sector basis elements
\begin{equation}
\label{gb71}
gb_{71}=\left(-12 gb_2+4 gb_{13}+32 gb_{16}+48 gb_{41}+36 gb_{42}-45 gb_{43}\right)/30 \, ,
\end{equation} 
while from the first one we obtain a relation between lower sector boundaries. We comment here that, as it was expected, in the relations between the boundaries the prefactors are just numbers because we have a pure U.T. basis.

From this method, we can also obtain other kinds of relations. Here comes our second example. Consider the master integrals of the same sector, $F_{41}$ and $F_{42}$, for which expansion-by-regions yields the regions $x \rightarrow 0: x^{-3 \varepsilon}, \, x^0$ and $x \rightarrow 0: x^{-1-3\varepsilon}, \, x^{-3 \varepsilon}, \, x^0$, respectively. Going back to the resummation matrix for these integrals we find two additional regions which we set to zero but from both, we find the same relation
\begin{equation}
\label{gb41one}
gb_{42}=-gb_2/2-gb_{13}/4-2 gb_{16}-3gb_{41} \, ,
\end{equation}
thus we can not express both of the boundaries of this sector in terms of lower boundaries. But we can obtain relations between boundaries and asymptotic limits by setting equal the result of the resummation matrix with that of the expansion-by-regions method, providing relations of the form
\begin{equation}
\label{gb41two}
gb_{41}= F_{41}^{\text{soft}} s_{12} \varepsilon^5 +gb_{2}/9-gb_{13}/12-2gb_{16}/3 \, ,
\end{equation}
where $F_{41}^{\text{soft}}$ is the value of the master integral $F_{42}$ in the region $x^{-3\varepsilon}$. Thus using the above relation one can compute the boundaries of this sector by computing only $F_{41}^{\text{soft}}$. Relations of the form such as (\ref{gb71}) and (\ref{gb41two}) are the two possible kinds of relations that we can obtain using this method.

By applying this method for all the basis elements we obtain in general $41$ relations of whom the $28$ are pure relations between boundaries (of the form (\ref{gb71})). Thus the problem of computing 41 boundaries is reduced to the calculation of the following $13$ asymptotic regions of master integrals
\begin{equation*}
\{ F_8^{\text{hard}}, \, F_9^{\text{hard}}, \, F_{12}^{\text{hard}}, \, F_{13}^{\text{hard}}, \, F_{16}^{\text{hard}}, F_{20}^{\text{hard}}, \, F_{27}^{\text{hard}}, \, F_{29}^{\text{hard}}, F_{32}^{\text{soft}}, \, F_{39}^{\text{soft}}, \, F_{41}^{\text{soft}}, \, F_{51}^{\text{hard}}, \, F_{56}^{\text{hard}} \},
\end{equation*}
where with "hard" we denote the contribution from the $x^0$ region and with "soft" from the $x^{-3\varepsilon}$. From the above set of asymptotic limits it is clear that there is no need to compute boundary conditions for high sectors, which in general is difficult. These contributions from the asymptotic limits that are left are easy to be calculated and we will describe in the next subsection how we did so.

\subsection{Contributing regions of master integrals}

\begin{figure} [h!]
\centering
\includegraphics[width=5 in]{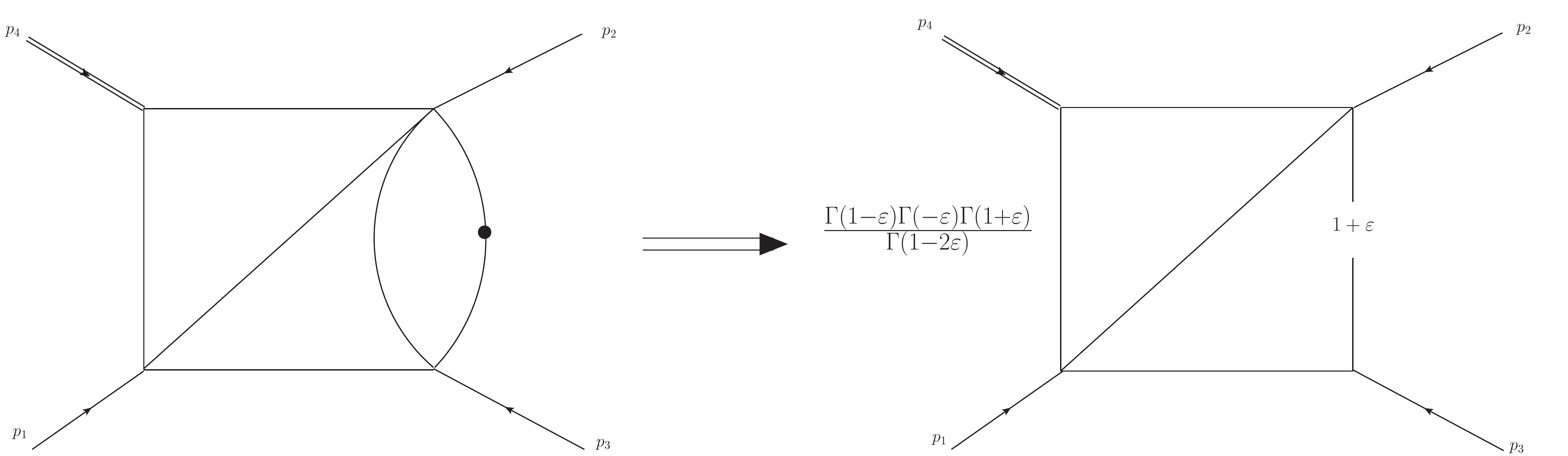}
\caption{Integrating out in $F_{32}$ the bubble subintegral we obtain a two-loop integral with 5 propagators, where the one related with the bubble subintegral has its power shifted by $\varepsilon$.}
\end{figure}

From the set of the asymptotic limits the calculation of the hard limits was performed in the momentum space with the use of the method of expansion-by-regions and IBP reduction. The SDE parametrization of the propagators makes it significantly easier in this case \cite{SDE1,SDE2,SDE3}. In fact, the hard asymptotic limits are equal to some of the known MI (subsection 3.1) and thus plugging them back into the relations between boundaries and asymptotic limits we found some extra relations between boundaries.

Regarding the soft limits, we employed a strategy depending on the Feynman - parameter representation of the integrals under consideration, as well as a technique of integrating out bubble subintegrals inspired by \cite{Henn:2013tua}. For the case in point, the three integrals whose regions we want to compute, contain bubble subintegrals, i.e. each one can be written as a two-loop integral with a bubble insertion. As explained in \cite{Henn:2013tua}, one can always integrate out bubble subintegrals and obtain a lower loop integral with some powers shifted by $\varepsilon$. More specifically we have
\begin{equation}\label{eq:bubble}
\int \frac{d^d k}{i \pi^{d/2}} \frac{1}{(k^2)^{a_1}((k+p)^2)^{a_2}}=\frac{\Gamma(a-d/2)\Gamma(d/2-a_1)\Gamma(d/2-a2_)}{\Gamma(a_1)\Gamma(a_2)\Gamma(d-a)} (p^2)^{d/2-a}   
\end{equation}
where $a=a_1+a_2$. For the cases that we are interested in, $a_1 = 2$ and $a_2 = 1$, and according to \cite{Henn:2013tua}, after integrating out the bubble subintegral we arrive at a two-loop integral with one index shifted from 1 to $1+\varepsilon$.

In the following, we explain in detail the computation of the required region for $F_{32}$. The remaining regions can be computed in a similar manner. According to our notation
\begin{equation}
F_{32}=\int \frac{d^dk_1 d^d k_2 d^d k_3}{(i\pi^{d/2})^3} \frac{e^{3\varepsilon\gamma_E}}{k_2^2 (k_1+k_2)^2 (k_2+k_3)^2 (k_1+p_{12})^2 (k_1+x p_1)^2 (k_3+p_{123})^4}
\end{equation}
The bubble subintegral is
\begin{equation}
\int \frac{d^d k_3}{i \pi^{d/2}} \frac{1}{(k_2+k_3)^2 (k_3+p_{123})^4}
\end{equation}
By making the shift $k_3 \to k -p_{123}$ we may write it as in \eqref{eq:bubble}. What is left is a two-loop integral with five propagators (see \textbf{Figure 3}). The departure from the usual case is the fact that one propagator is raised to the power of $1+\varepsilon$. If one tries to find the regions contributing to the $x\to 0$ limit of this integral we see that there are two regions, namely $(x^0, x^{-3\varepsilon})$ as for the full three-loop integral. We are interested in the $x^{-3\varepsilon}$ region, which now is significantly less complicated than the one resulting from the full three-loop integral analysis. After performing the necessary integrations, we may assemble the final result by multiplying the result of the bubble integration, namely \eqref{eq:bubble} for $a_1 = 2$ and $a_2 = 1$, with the result of the two-loop region and expand up to the desired power in the dimensional regulator.


\subsection{Application to other families}

To point out the significance of the method described in subsection \ref{3.2} we briefly mention applications of it to other families. With regards to non-planar topologies, using this method for the two non-planar families of the doublebox with one external mass we were able to obtain relations for the boundaries of the non-planar basis elements in terms of the known planar ones \cite{Syrrakosthesis}. More interestingly, regarding problems with many scales, this method was successfully used for the calculation of the boundaries of the three planar two-loop five-point families with one external mass \cite{5box}, where it produced 47, 42 and 61 pure relations between boundaries for the P1, P2 and P3 family, respectively. 

\section{Massless three-loop ladder-box}
\label{sec:massless}

It is interesting to see how we can extract the UT basis and solution for the massless three-loop ladder-box using our results from the massive one. An important feature of the SDE approach is that by correctly taking the $x\to 1$ limit \cite{SDE3, Papadopoulos:2019iam} of the solution of a specific integral family with $n$ scales, one can arrive at the solution of the corresponding family with $n-1$ scales. In this particular case, taking the $x \to 1$ limit of the one-mass triple box will yield the solution for the massless one. 

\subsection{\boldmath The $x\to 1$ limit}
 Taking the $x\to 1$ limit of the one-mass three-loop ladder-box amounts to performing the following manipulations. First, we rewrite our solution as an expansion in $\log(1-x)$ in the following form \cite{SDE3}
\begin{equation}\label{eq:solexp1}
    \textbf{g}=\sum_{n\geq 0} \epsilon^{n} \sum_{i=0}^{n}\frac{1}{i!}\textbf{c}^{(n)}_{i} \log^i(1-x)
\end{equation}
where all coefficients $\textbf{c}^{(n)}_{i}$ are finite in the limit $x\to 1$. This can be straightforwardly achieved, starting from the original solution \eqref{solution} and transporting all letters $l=1$ of GPLs to the right, according to their known shuffle properties. Having done that, we may define the regular part of $\textbf{g}$ at $x=1$ as
\begin{equation}\label{eq:reg1}
   \textbf{g}_{reg}=\sum_{n\geq 0}\epsilon^n \textbf{c}^{(n)}_{0}
\end{equation}
and through $\textbf{g}_{reg}$, the truncated part of $\textbf{g}$ as \cite{SDE3}
\begin{equation}\label{eq:trunc}
    \textbf{g}_{trunc} = \textbf{g}_{reg}(x=1)
\end{equation}
If we return to the canonical DE \eqref{de} and single out the part singular at $x=1$, then the full solution of the problem can be written schematically as follows
\begin{equation}\label{eq:solexp2}
    \textbf{g}= e^{\epsilon \textbf{M}_1 \log(1-x)} \textbf{g}_{reg}
\end{equation}
Because $\textbf{M}_1$ is by definition a square matrix, we can always find its Jordan matrix decomposition
\begin{equation}\label{eq:jordan1}
    \textbf{M}_1 = \textbf{S}_1 \textbf{D}_1 \textbf{S}_{1}^{-1}
\end{equation}
We may then define the \textbf{resummation matrix} $\textbf{R}_1$ as follows
\begin{equation}\label{eq:resum1}
    \textbf{R}_1 = e^{\epsilon \textbf{M}_1 \log(1-x)} = \textbf{S}_1 e^{\epsilon \textbf{D}_1 \log(1-x)} \textbf{S}_{1}^{-1}
\end{equation}
It is clear from \eqref{eq:solexp2} that when \eqref{eq:resum1} acts on \eqref{eq:reg1} we get the full solution of the problem. The resummation matrix $\textbf{R}_1$ has terms $(1-x)^{a_{i} \epsilon}$, with $a_{i}$ the eigenvalues of $\textbf{M}_1$. Setting these terms equal to zero results in a purely numerical matrix $\textbf{R}_{10}$
\begin{equation}
    \textbf{R}_1 \rightarrow \textbf{R}_{10}
\end{equation}
Now, finding the $x\to1$ limit of the original solution \eqref{solution} amounts to $\textbf{R}_{10}$ acting on $\textbf{g}_{trunc}$.
\begin{equation}\label{eq:glim1}
    \textbf{g}_{x\to 1} = \textbf{R}_{10} \textbf{g}_{trunc}
\end{equation}

\subsection{From massive to massless}
Remember that our goal is to find a UT basis for the massless three-loop ladder-box. From the massive one, the result of $$\lim_{x\to\ 1} \textbf{g}(x)$$ is an array of 83 UT elements. If we look at this result from the perspective of the massless three-loop ladder-box, some of these UT elements will be the 26 Basis Elements (BE) for the massless three-loop ladder-box and the rest will be reducible to the 26 massless BE or zero. We may distinguish them using IBP relations but first we may exploit an interesting feature of the $\textbf{R}_{10}$ matrix in order to simplify the procedure.

It turns out that $\textbf{R}_{10}$ is an \textit{idempotent} matrix. Idempotent matrices have the following properties, all of which are satisfied by $\textbf{R}_{10}$:
\begin{enumerate}
    \item $\textbf{X} = \textbf{X}^2$
    \item singular except the identity matrix $\textbf{I}$
    \item eigenvalues of $\textbf{X} = 0,1$
    \item Trace$\big(\textbf{X}\big)=$ Rank$\big(\textbf{X}\big)$
    \item $\textbf{I}-\textbf{X}$ also idempotent
\end{enumerate}
Since $ \textbf{R}_{10} = \textbf{R}_{10}^2$, if we act with $\textbf{R}_{10}$ on $\textbf{g}_{x\to 1}$ yields
\begin{align}\label{eq:methodx1}
    \textbf{R}_{10} \textbf{g}_{x\to 1} &= \textbf{R}_{10}^2 \textbf{g}_{trunc} \nonumber \\
    &=\textbf{R}_{10} \textbf{g}_{trunc} \nonumber \\
    &=\textbf{g}_{x\to 1}
\end{align}
This relation, solved as an equation for each row, will yield the relations between the reducible UT elements and the BE as well as those who are equal to zero. We found 36 relations, which is expected since the rank of the purely numerical resummation matrix is 36, e.g.\footnote{In the following we use again the notation $gb_i$ to indicate the $x\to 1$ limit of the corresponding $g_i$.}
\begin{align}
\text{gb}_{24} &= 0, \\
\text{gb}_{49} &= \frac{1}{2} (2 \text{gb}_{35}-\text{gb}_{33}), \\
\text{gb}_{80} &= \frac{1}{36} (4 \text{gb}_{1}-6 \text{gb}_{4}-48 \text{gb}_{5}+192 \text{gb}_{10}+128 \text{gb}_{14}-72 \text{gb}_{15}-96 \text{gb}_{22} \nonumber \\
&-12 \text{gb}_{33}-192 \text{gb}_{36}+3 \text{gb}_{40}-12 \text{gb}_{42}-9 \text{gb}_{43}-8 \text{gb}_{48}-24 \text{gb}_{53} \nonumber \\
&-12 \text{gb}_{55}+36 \text{gb}_{72}-6 \text{gb}_{75}-108 \text{gb}_{76}-12 \text{gb}_{77})
\end{align}

The resulting relations can be verified in two ways.
\begin{enumerate}
    \item Using IBP relations.
    \item Using the analytic expressions for the $x\to 1$ limit. 
\end{enumerate}
These 36 relations leave 47 UT elements as potential candidates for the 26 BE of the massless three-loop ladder-box, namely
\begin{align*}
    \{&\text{gb}_1,\text{gb}_2,\text{gb}_4,\text{gb}_5,\text{gb}_6,\text{gb}_7,\text{gb}_{10},\text{gb}_{14},\text{gb}_{15},\text{gb}_{17},\text{gb}_{18},\text{gb}_{19},\text{gb}_{22},\text{gb}_{30},\text{gb}_{31},\text{gb}_{33},\text{gb}_{35}, \nonumber \\
    &\text{gb}_{36},\text{gb}_{37},\text{gb}_{38},\text{gb}_{40},\text{gb}_{42},\text{gb}_{43},\text{gb}_{44},\text{gb}_{46},\text{gb}_{48},\text{gb}_{53},\text{gb}_{55},\text{gb}_{58},\text{gb}_{59},\text{gb}_{62},\text{gb}_{63},\text{gb}_{64},\nonumber \\
    &\text{gb}_{65},\text{gb}_{66},\text{gb}_{67},\text{gb}_{68},\text{gb}_{69},\text{gb}_{70},\text{gb}_{71},\text{gb}_{72},\text{gb}_{75},\text{gb}_{76},\text{gb}_{77},\text{gb}_{81},\text{gb}_{82},\text{gb}_{83}\}
\end{align*}
These $\text{gb}_i$ correspond to the $g_i$ of \textbf{Table 1} and \textbf{Table 2} in the limit $x\to 1$. To distinguish among them we may perform an IBP reduction. Since we want to extract the BE for the massless three-loop ladder-box from the BE of the massive one, we first substitute each $\text{gb}_i$ with its corresponding $g_i$ of Tables 1 and 2, and we set $x=1$ explicitly. We now have a set of 47 UT elements written in terms of certain FI. Via IBP reduction, we may reduce these FI to a set of MI for the massless three-loop ladder-box, and we can see that there are 26 linearly independent UT elements. 


\subsection{UT basis for the massless three-loop ladder-box}
The resulting UT basis for the massless three-loop ladder-box in terms of the ones from the massive three-loop ladder-box are 
\begin{align*}
    \{&\text{gb}_1,\text{gb}_2,\text{gb}_4,\text{gb}_5,\text{gb}_6,\text{gb}_7,\text{gb}_{17},\text{gb}_{18},\text{gb}_{19},\text{gb}_{30},\text{gb}_{31},\text{gb}_{35},\text{gb}_{36},\text{gb}_{37},\text{gb}_{44},\text{gb}_{53}, \nonumber \\ &\text{gb}_{55},\text{gb}_{62}, \text{gb}_{63},\text{gb}_{64},\text{gb}_{65},\text{gb}_{68},\text{gb}_{69},\text{gb}_{81},\text{gb}_{82},\text{gb}_{83}\}
\end{align*}
which can be written in terms of 26 MI as follows
\begin{align}
&\text{g}_{1} =  s_{12} \varepsilon^3 G_{0,0,2,2,2,0,1,0,0,0,0,0,0,0,0}, \nonumber \\
&\text{g}_{2} =  s_{23} \varepsilon^3 G_{0,0,0,2,2,1,0,0,0,2,0,0,0,0,0}, \nonumber \\
&\text{g}_{3} =  s_{12}^2 \varepsilon^3 G_{0,2,2,1,0,0,2,0,1,0,0,0,0,0,0}, \nonumber \\
&\text{g}_{4} =  s_{12} \varepsilon^3 (2 \varepsilon +1) G_{0,1,0,1,1,0,2,0,2,0,0,0,0,0,0}, \nonumber \\
&\text{g}_{5} =  s_{12} \varepsilon^4 G_{1,0,0,2,1,2,1,0,0,0,0,0,0,0,0}, \nonumber \\
&\text{g}_{6} =  s_{12} \varepsilon^4 G_{0,1,0,2,1,2,1,0,0,0,0,0,0,0,0}, \nonumber \\
&\text{g}_{7} =  s_{12}^3 \varepsilon^3 G_{2,2,2,0,0,0,1,1,1,0,0,0,0,0,0}, \nonumber \\
&\text{g}_{8} =  s_{12} (1-2 \varepsilon ) \varepsilon^4 G_{1,0,1,2,1,0,1,0,1,0,0,0,0,0,0}, \nonumber \\
&\text{g}_{9} =  s_{12}^2 \varepsilon^4 G_{2,1,0,0,1,2,1,1,0,0,0,0,0,0,0}, \nonumber \\
&\text{g}_{10} =  s_{12} s_{23} \varepsilon^4 G_{1,0,0,2,1,2,1,0,0,1,0,0,0,0,0}, \nonumber \\
&\text{g}_{11} =  \varepsilon^5 (-s_{12}-s_{23}) G_{0,1,0,1,1,2,1,0,0,1,0,0,0,0,0}, \nonumber \\
&\text{g}_{12} =  s_{12} s_{23} \varepsilon^4 G_{0,1,0,2,1,2,0,1,0,1,0,0,0,0,0}, \nonumber \\
&\text{g}_{13} =  \frac{1}{4} s_{12} \varepsilon^3 \left(\frac{4 \varepsilon  (2 \varepsilon -1) G_{0,1,0,2,1,2,-1,1,0,1,0,0,0,0,0}}{\varepsilon -1}+G_{0,0,0,2,2,1,0,0,0,2,0,0,0,0,0}\right), \nonumber \\
&\text{g}_{14} =  \varepsilon^5 (-s_{12}-s_{23}) G_{0,0,1,1,2,1,1,0,0,1,0,0,0,0,0}, \nonumber \\
&\text{g}_{15} =  s_{12} \varepsilon^6 G_{1,0,1,1,1,1,1,1,0,0,0,0,0,0,0}, \nonumber \\
&\text{g}_{16} =  -s_{12}^2 \varepsilon^5 G_{1,0,1,1,1,1,0,2,0,1,0,0,0,0,0}, \nonumber \\
&\text{g}_{17} =  s_{12} s_{23} \varepsilon^5 G_{1,0,1,2,1,1,0,1,0,1,0,0,0,0,0}, \nonumber \\
&\text{g}_{18} =  s_{12}^2 s_{23} \varepsilon^5 G_{1,1,0,1,1,2,1,1,0,1,0,0,0,0,0}, \nonumber \\
&\text{g}_{19} =  s_{12}^2 \varepsilon^5 (2 \varepsilon -1) G_{1,1,0,1,1,1,1,1,0,1,0,0,0,0,0}, \nonumber \\
&\text{g}_{20} =  -s_{12} \varepsilon^6 (s_{12}+s_{23}) G_{1,0,1,1,1,1,1,1,0,1,0,0,0,0,0}, \nonumber \\
&\text{g}_{21} =  s_{12}^2 s_{23} \varepsilon^5 G_{1,0,1,1,2,1,1,1,0,1,0,0,0,0,0}, \nonumber \\
&\text{g}_{22} =  s_{12}^2 \varepsilon^5 (2 \varepsilon -1) G_{1,0,1,1,1,1,1,0,1,1,0,0,0,0,0}, \nonumber \\
&\text{g}_{23} =  s_{12}^2 s_{23} \varepsilon^5 G_{1,0,1,1,2,1,1,0,1,1,0,0,0,0,0}, \nonumber \\
&\text{g}_{24} =  s_{12}^3 s_{23} \varepsilon^6 G_{1,1,1,1,1,1,1,1,1,1,0,0,0,0,0}, \nonumber \\
&\text{g}_{25} =  -s_{12}^3 \varepsilon^6 G_{1,1,1,1,1,1,1,1,1,1,-1,0,0,0,0}, \nonumber \\
&\text{g}_{26} =  -s_{12}^3 \varepsilon^6 G_{1,1,1,1,1,1,1,1,1,1,0,-1,0,0,0}
\end{align}
The chosen normalisation of the FI is
\begin{equation}
    G_{a_{1}, \ldots, a_{15}}\left(\left\{p_{j}\right\}, \varepsilon\right)=(-s_{12})^{3\varepsilon}\int\left(\prod_{l=1}^{3} \frac{d^{d} k_{l}}{i \pi^{d / 2}}\right) \frac{e^{3 \varepsilon \gamma_{E}}}{D_{1}^{a_{1}} \ldots D_{15}^{a_{15}}} \quad \text { with } \quad d=4-2 \varepsilon
\end{equation}
with the propagators being the $x\to 1$ limit of \eqref{SDEprop}
\begin{equation}
\begin{array}{l}
D_{1}=\left(k_{1}+p_{12}\right)^{2}, \quad D_{2}=\left(k_{2}-p_{12}\right)^{2}, \quad D_{3}=\left(k_{3}+p_{12}\right)^{2}, \quad D_{4}=\left(k_{1}+k_{2}\right)^{2} \\
D_{5}=\left(k_{2}+k_{3}\right)^{2}, \quad D_{6}=\left(k_{3}+p_{123}\right)^{2}, \quad D_{7}=k_{1}^{2}, \quad D_{8}=k_{2}^{2}, \quad D_{9}=k_{3}^{2} \\
D_{10}=\left(k_{1}+p_{1}\right)^{2}, \quad D_{11}=\left(k_{1}+p_{123}\right)^{2}, \quad D_{12}=\left(k_{2}-p_{123}\right)^{2} \\
D_{13}=\left(k_{2}-p_{1}\right)^{2}, \quad D_{14}=\left(k_{3}+p_{1}\right)^{2}, \quad \text{and} \quad D_{15}=\left(k_{1}-k_{3}\right)^{2}
\end{array}
\end{equation}

\subsection{Checks performed}
Starting from the $x\to 1$ limit of our result for the massive three-loop ladder-box, we used \texttt{HyperInt} \cite{hyper} to write our expressions in terms of GPLs, with indices from the set $\{0,1\}$ and argument the variable $y = s_{23}/s_{12}$. We compared analytically our results for the three top sector basis elements with the ones given by \cite{Henn:2013tua} and numerically for all basis elements with \texttt{pySecDec} \cite{pysecdec} in the Euclidean region. In both cases perfect agreement was found. The results and the canonical basis of the massless case can be found in the ancillary files.

\subsection{Applications in other families of MI and optimization}

The above method has been applied to the families of both planar and non-planar two-loop box MI with one \cite{Syrrakosthesis} and with two external masses \cite{Henn2, SDE2, Papadopoulos:2019iam}, as well as to the two-loop planar pentabox families with one external mass. Within the SDE approach, we have a freedom of choosing the mapping of the original external momenta to the $x$-parameterisation. So far we have used two different parametrisations, with one $x$ (in this paper and for the two-loop box with one external mass), and with two $x$'s\footnote{By one and two $x$'s we mean that in the first case, only one external momentum is parametrized as  $x p_i$, whereas in the latter case we have two external momenta parametrized  as  $x p_i$.}, in \cite{SDE2,SDE3}. Our results so far indicate that the choice of two $x$'s is the most optimal for the application of this method, since after defining the resummation matrix and obtaining its purely numerical form, \eqref{eq:methodx1} yields a number of linearly independent UT elements exactly equal to the number of MI of the problem with one scale less, i.e. we do not need to perform an IBP reduction to determine the new Basis Elements. This can also be seen by computing the rank of the numerical resummation matrix. Exploiting this fact drastically simplifies the process of extracting a canonical basis for the problem with one scale less, however it is still yet not clear why the two parametrizations yield so different results.

\section{Conclusions}
\label{sec:conlusions}

In the last decade we have witnessed an explosion of new and important results in the field of  analytical calculations of MI using the method of DE. This huge push forward has been facilitated mainly by the introduction of the \textit{canonical form} of the DE \cite{Henn1}, as well as a better understanding of the class of functions that arise from the solution of the aforementioned DE \cite{Duhr}. However, as problems become more complicated the development of new methods and ideas for solving DE \cite{Moriello1, Moriello2, Moriello3, CB9, Hidding} and more general for making computations within QCD \cite{Anastasiou1,Anastasiou2,Papadopoulos:2019iam,Heinrich} becomes imperative, since current methods come under extreme pressure. 

In this paper we presented resummation methods which simplify the process of solving a canonical DE, as well as the extraction of the solution of a problem with $n-1$ scales, once the solution for the corresponding problem with $n$ scales is known. These methods are tailored to be used with the SDE approach, and although presented here within the problem of the three-loop ladder-box with one external mass they have been successfully applied to a variety of other problems as well, proven to work independently of the topology (planar or non-planar) the number of loops (two or three) and the number of scales of the family. The successful application of these methods gives us confidence that their use can indeed benefit the effort of providing analytical solutions for MI through the use of canonical DE, in conjunction with the SDE approach.

As a next step we see the application of these methods for obtaining analytic solutions for other families of phenomenological interest such as the non-planar two-loop five-point families with one external mass and the rest of the three-loop four-point families with one external mass. Moreover, of great interest would be the application of the proposed method for obtaining boundary terms to the standard method of DE where more than one integrating variables are present, as well as their automation in the form of \texttt{Mathematica} packages.

The precision frontier, initiated by advances from our experimental colleagues \cite{Heinrich,Amoroso:2020lgh}, is a major challenge for contemporary Particle Physics. We hope that our contribution will provide theorists with new tools to meet this challenge. 

\acknowledgments
\begin{wrapfigure}{r}{0.35\textwidth}
    \centering
    \includegraphics[width=0.5\textwidth]{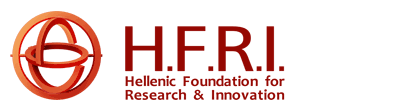}
\end{wrapfigure}
We would like to thank Costas Papadopoulos and Chris Wever for useful discussions and suggestions on the draft. The research work was supported by the Hellenic Foundation for Research and Innovation (HFRI) under the HFRI PhD Fellowship grant (Fellowship Number: 554).

\begin{table}[h!]
\begin{adjustwidth}{-1.in}{-1.in} 
\begin{center}
\begin{tabular}{| c | c |} 
 \hline
 \textbf{Basis Element} & \textbf{Asymptotic Limit of Master Integral $x \rightarrow 0$} \\
 \hline
 $g_1\equiv s_{12} \varepsilon^3 F_{1}$ & $F_1 \equiv G_{0, 0, 2, 2, 2, 0, 1, 0, 0, 0, 0, 0, 0, 0, 0} \sim \text{closed form}$  \\ 
 \hline
 $g_2 \equiv s_{23} x \varepsilon^3 F_{2}$ & $F_2 \equiv G_{0, 0, 0, 2, 2, 1, 0, 0, 0, 2, 0, 0, 0, 0, 0} \sim \text{closed form}$  \\
 \hline
 $g_3 \equiv s_{12} (1-x) \varepsilon^3 F_{3}$ & $F_{3} \equiv G_{0, 0, 2, 2, 2, 0, 0, 0, 0, 1, 0, 0, 0, 0,0} \sim \text{closed form}$  \\ 
 \hline
 $g_4 \equiv s_{12}^2 \varepsilon^3 F_{4}$ & $F_{4} \equiv G_{0, 2, 2, 1, 0, 0, 2, 0, 1, 0, 0, 0, 0, 0, 0} \sim \text{closed form}$   \\
 \hline
 $g_5 \equiv s_{12} \varepsilon^3 (1 + 2 \varepsilon) F_{5}$ & $F_{5}\equiv G_{0, 1, 0, 1, 1, 0, 2, 0, 2, 0, 0, 0, 0, 0, 0} \sim \text{closed form}$   \\ 
 \hline
 $g_6 \equiv s_{12} \varepsilon^4 F_{6}$ & $F_{6} \equiv G_{1, 0, 0, 2, 1, 2, 1, 0, 0, 0, 0, 0, 0, 0, 0} \sim \text{closed form}$  \\
 \hline
 $g_7 \equiv s_{12} \varepsilon^4 F_{7}$ & $F_{7} \equiv G_{0, 1, 0, 2, 1, 2, 1, 0, 0, 0, 0, 0, 0, 0, 0} \sim \text{closed form}$  \\ 
 \hline
 $g_8 \equiv s_{12}^2 (1 - x) \varepsilon^3 F_{8}$ & $F_{8} \equiv G_{0, 2, 2, 1, 0, 0, 0, 0, 1, 2, 0, 0, 0, 0, 0} \sim x^0$   \\
 \hline
 $g_9 \equiv s_{12}^2 (1 - x) \varepsilon^3 F_{9}$ & $F_{9} \equiv G_{2, 0, 2, 0, 1, 0, 0, 2, 0, 1, 0, 0, 0, 0, 0} \sim x^0$   \\ 
 \hline
 $g_{10} \equiv -s_{12} x \varepsilon^4 F_{10}$ & $F_{10} \equiv G_{0, 0, 2, 2, 1, 0, 0, 1, 0, 1, 0, 0, 0, 0, 0} \sim x^0$   \\
 \hline
 $g_{11} \equiv -s_{12} x \varepsilon^4 F_{11}$ & $F_{11} \equiv G_{1, 0, 0, 2, 1, 0, 0, 0, 2, 1, 0, 0, 0, 0, 0} \sim x^0$   \\ 
 \hline
 $g_{12} \equiv (s_{12}(1 - x) - s_{23} x) \varepsilon^4 F_{12}$ & $F_{12} \equiv G_{1, 0, 0, 2, 1, 2, 0, 0, 0, 1, 0, 0, 0, 0, 0} \sim x^0, \, x^{-3\varepsilon}$   \\
 \hline
 $g_{13} \equiv s_{12} (1 - x) \varepsilon^3 (s_{12} F_{13}-4 \varepsilon F_{14})$ & $F_{13} \equiv G_{0, 2, 0, 2, 1, 0, 0, 0, 2, 1, 0, 0, 0, 0, 0} \sim x^0$   \\ 
 \hline
 $g_{14} \equiv -s_{12} x \varepsilon^4 F_{14}$ & $F_{14} \equiv G_{0, 1, 0, 2, 1, 0, 0, 0, 2, 1, 0, 0, 0, 0, 0} \sim x^0$    \\
 \hline
 $g_{15} \equiv -s_{12} x \varepsilon^4 F_{15}$ & $F_{15} \equiv G_{0, 0, 1, 1, 2, 0, 0, 0, 1, 2, 0, 0, 0, 0, 0} \sim x^0$   \\ 
 \hline
 $g_{16} \equiv (s_{12} (1-x)-s_{23} x) \varepsilon^4 F_{16} $ & $F_{16} \equiv G_{0, 1, 0, 2, 1, 2, 0, 0, 0, 1, 0, 0, 0, 0, 0} \sim x^0, \, x^{-3\varepsilon}$    \\
 \hline
 $g_{17} \equiv s_{12}^3 \varepsilon^3 F_{17}$ & $F_{17} \equiv G_{2, 2, 2, 0, 0, 0, 1, 1, 1, 0, 0, 0, 0, 0, 0} \sim \text{closed form}$    \\
 \hline
 $g_{18} \equiv s_{12} (1-2 \varepsilon) \varepsilon^4 F_{18}$ & $F_{18} \equiv G_{1, 0, 1, 2, 1, 0, 1, 0, 1, 0, 0, 0, 0, 0, 0} \sim \text{closed form}$    \\ 
 \hline
 $g_{19} \equiv s_{12}^2 \varepsilon^4 F_{19}$ & $F_{19} \equiv G_{2, 1, 0, 0, 1, 2, 1, 1, 0, 0, 0, 0, 0, 0, 0} \sim \text{closed form}$   \\
 \hline
 $g_{20} \equiv s_{12}^3 (1-x) \varepsilon^3 F_{20}$ & $F_{20} \equiv G_{2, 2, 2, 0, 0, 0, 0, 1, 1, 1, 0, 0, 0, 0, 0} \sim x^0$    \\ 
 \hline
 $g_{21} \equiv -s_{12}^2 x \varepsilon^4 F_{21}$ & $F_{21} \equiv G_{1, 0, 2, 1, 0, 0, 0, 2, 1, 1, 0, 0, 0, 0, 0} \sim x^0$    \\
 \hline
 $g_{22} \equiv -s_{12}^2 x \varepsilon^4 F_{22}$ & $F_{22} \equiv G_{0, 1, 2, 2, 0, 0, 0, 1, 1, 1, 0, 0, 0, 0, 0} \sim x^0$   \\ 
 \hline
 $g_{23} \equiv -s_{12} x \varepsilon^5 F_{23}$ & $F_{23} \equiv G_{1, 0, 2, 1, 1, 0, 0, 1, 0, 1, 0, 0, 0, 0, 0} \sim x^0$   \\
 \hline
 $g_{24} \equiv s_{12}^2 (x-1) x \varepsilon^4 F_{24}$ & $F_{24} \equiv G_{1, 0, 2, 1, 1, 0, 0, 1, 0, 2, 0, 0, 0, 0, 0} \sim x^0$   \\
 \hline
 $g_{25} \equiv -s_{12} x \varepsilon^5 F_{25}$ & $F_{25} \equiv G_{1, 1, 0, 1, 1, 0, 0, 0, 2, 1, 0, 0, 0, 0, 0} \sim x^0$   \\ 
 \hline
 $g_{26} \equiv -s_{12} x \varepsilon^5 F_{26}$ & $F_{26} \equiv G_{1, 0, 1, 1, 2, 0, 0, 0, 1, 1, 0, 0, 0, 0, 0} \sim x^0$  \\
 \hline
 $g_{27} \equiv s_{12} (1 - x) \varepsilon^3 (s_{12} (1 + 2 \varepsilon) F_{27}-12 \varepsilon^2 F_{26})$ & $F_{27} \equiv G_{1, 0, 2, 1, 2, 0, 0, 0, 1, 1, 0, 0, 0, 0, 0} \sim x^0$   \\ 
 \hline
 $g_{28} \equiv -s_{12} x \varepsilon^5 F_{28}$ & $F_{28} \equiv G_{0, 1, 1, 2, 1, 0, 0, 0, 1, 1, 0, 0, 0, 0, 0} \sim x^0$   \\
 \hline
 $g_{29} \equiv -s_{12}^2 (x-1) \varepsilon^4 F_{29}$ & $F_{29} \equiv G_{2, 1, 0, 0, 1, 2, 0, 1, 0, 1, 0, 0, 0, 0, 0} \sim x^0$   \\ 
 \hline
 $g_{30} \equiv s_{12} s_{23} x \varepsilon^4 F_{30}$ & $F_{30} \equiv G_{1, 0, 0, 2, 1, 2, 1, 0, 0, 1, 0, 0, 0, 0, 0} \sim x^{-1-3\varepsilon}, \, x^{-3\varepsilon}, \, x^0$   \\
 \hline
 $g_{31} \equiv -(s_{12} + s_{23}) x \varepsilon^5 F_{31}$ & $F_{31} \equiv G_{0, 1, 0, 1, 1, 2, 1, 0, 0, 1, 0, 0, 0, 0, 0} \sim x^{-3\varepsilon}, \, x^0$   \\
 \hline
 $g_{32} \equiv (s_{12} + s_{23} x) \varepsilon^5 F_{32}$ & $F_{32} \equiv G_{1, 0, 0, 1, 1, 2, 0, 1, 0, 1, 0, 0, 0, 0, 0} \sim x^{-3\varepsilon}, \, x^0$   \\ 
 \hline
 $g_{33} \equiv s_{12} s_{23} x \varepsilon^4 F_{33}$ & $F_{33} \equiv G_{1, 0, 0, 2, 1, 2, 0, 1, 0, 1, 0, 0, 0, 0, 0} \sim x^{-1-3\varepsilon}, \, x^{-3\varepsilon}, \, x^0$   \\
 \hline
 $g_{34} \equiv s_{12}^2 (1-x) \varepsilon^4 F_{34}$ & $F_{34} \equiv G_{2, 0, 0, 1, 1, 2, 0, 1, 0, 1, 0, 0, 0, 0, 0} \sim x^{-3\varepsilon}, \, x^0$   \\ 
 \hline
 $g_{35} \equiv s_{12} s_{23} x \varepsilon^4 F_{35}$ & $F_{35} \equiv G_{0, 1, 0, 2, 1, 2, 0, 1, 0, 1, 0, 0, 0, 0, 0} \sim x^{-1-3\varepsilon}, \, x^{-3\varepsilon}, \, x^0$   \\
 \hline
 $g_{36} \equiv s_{12} \varepsilon^4 ((2 \varepsilon-1) F_{36} - (\varepsilon-1) F_{16})/(\varepsilon-1)$ & $F_{36} \equiv  G_{0, 1, 0, 2, 1, 2, -1, 1, 0, 1, 0, 0, 0, 0, 0}  \sim x^{-3\varepsilon}, \, x^0$   \\ 
 \hline
 $g_{37} \equiv -(s_{12} + s_{23}) x \varepsilon^5 F_{37}$ & $F_{37} \equiv G_{0, 0, 1, 1, 2, 1, 1, 0, 0, 1, 0, 0, 0, 0, 0} \sim x^{-3\varepsilon}, \, x^0$   \\
 \hline
 $g_{38} \equiv -(s_{12} + s_{23}) x \varepsilon^5 F_{38}$ & $F_{38} \equiv G_{0, 0, 1, 2, 1, 1, 0, 1, 0, 1, 0, 0, 0, 0, 0} \sim x^{-3\varepsilon}, \, x^0$   \\
 \hline
 $g_{39} \equiv (s_{12} + s_{23} x) \varepsilon^5 F_{39}$ & $F_{39} \equiv G_{1, 0, 0, 1, 2, 1, 0, 0, 1, 1, 0, 0, 0, 0, 0} \sim x^{-3\varepsilon}, \, x^0$   \\ 
 \hline
 $g_{40} \equiv s_{12} s_{23} x \varepsilon^4 F_{40}$ & $F_{40} \equiv G_{1, 0, 0, 2, 2, 1, 0, 0, 1, 1, 0, 0, 0, 0, 0} \sim x^{-1-3\varepsilon}, \, x^{-3\varepsilon}, \, x^0$   \\
 \hline
 $g_{41} \equiv (s_{12} + s_{23} x) \varepsilon^5 F_{41}$ & $F_{41} \equiv G_{0, 1, 0, 2, 1, 1, 0, 0, 1, 1, 0, 0, 0, 0, 0} \sim  x^{-3\varepsilon}, \, x^0$  \\ 
 \hline
 $g_{42} \equiv s_{12} s_{23} x \varepsilon^4 F_{42}$ & $F_{42} \equiv G_{0, 1, 0, 2, 2, 1, 0, 0, 1, 1, 0, 0, 0, 0, 0} \sim x^{-1-3\varepsilon}, \, x^{-3\varepsilon}, \, x^0$   \\
 \hline
\end{tabular}
\end{center}
\end{adjustwidth}
\caption{The First 42 U.T. basis elements of the basis presented in \cite{mastrolia} written in the SDE notation.}
\label{table:1}
\end{table}

\begin{table}[h!]
\begin{adjustwidth}{-1.in}{-1.in} 
\begin{center}
\begin{tabular}{| c | c |} 
 \hline
 \textbf{Basis Element} & \textbf{Asymptotic Limit of Master Integral $x \rightarrow 0$} \\
 \hline
 $g_{43} \equiv s_{12} s_{23} x \varepsilon^4 F_{43}$ & $F_{43} \equiv G_{0, 0, 1, 1, 2, 1, 0, 0, 1, 2, 0, 0, 0, 0, 0} \sim x^{-1-3\varepsilon}, \, x^{-3\varepsilon}, \, x^0 $  \\ 
 \hline
 $g_{44} \equiv s_{12} \varepsilon^6 F_{44}$ & $F_{44} \equiv G_{1, 0, 1, 1, 1, 1, 1, 1, 0, 0, 0, 0, 0, 0, 0} \sim \text{closed form}$  \\
 \hline
 $g_{45} \equiv -s_{12}^2 x \varepsilon^5 F_{45}$ & $F_{45} \equiv G_{1, 1, 2, 1, 0, 0, 0, 1, 1, 1, 0, 0, 0, 0, 0} \sim x^0$  \\ 
 \hline
 $g_{46} \equiv -s_{12} x \varepsilon^6 F_{46}$ & $F_{46} \equiv G_{0, 1, 1, 1, 1, 0, 1, 0, 1, 1, 0, 0, 0, 0, 0} \sim x^0$   \\
 \hline
 $g_{47} \equiv -s_{12} x \varepsilon^6 F_{47}$ & $F_{47} \equiv G_{1, 0, 1, 1, 1, 0, 0, 1, 1, 1, 0, 0, 0, 0, 0} \sim x^0$   \\ 
 \hline
 $g_{48} \equiv -s_{12}^2 x \varepsilon^4 (1 + \varepsilon) F_{48}$ & $F_{48} \equiv G_{1, 0, 1, 1, 1, 0, 0, 2, 1, 1, 0, 0, 0, 0, 0} \sim x^0$  \\
 \hline
 $g_{49} \equiv s_{12} (s_{12} (1-x) - s_{23} x) \varepsilon^5 F_{49}$ & $F_{49} \equiv G_{1, 1, 0, 1, 1, 2, 0, 1, 0, 1, 0, 0, 0, 0, 0} \sim x^{-3\varepsilon}, \, x^0$  \\ 
 \hline
 $g_{50} \equiv s_{12} x \varepsilon^5 (2 \varepsilon-1) F_{50}$ & $F_{50} \equiv G_{1, 1, 0, 1, 1, 1, 0, 1, 0, 1, 0, 0, 0, 0, 0} \sim x^0$   \\
 \hline
 $g_{51} \equiv (s_{12} (1-x) - s_{23} x) \varepsilon^6 F_{51}$ & $F_{51} \equiv G_{1, 0, 1, 1, 1, 1, 0, 1, 0, 1, 0, 0, 0, 0, 0} \sim x^0$   \\ 
 \hline
 $g_{52} \equiv s_{12}^2 (1-x) \varepsilon^5 F_{52}$ & $F_{52} \equiv G_{2, 0, 1, 1, 1, 1, 0, 1, 0, 1, 0, 0, 0, 0, 0} \sim x^0$   \\
 \hline
 $g_{53} \equiv -s_{12}^2 x \varepsilon^5 F_{53}$ & $F_{53} \equiv G_{1, 0, 1, 1, 1, 1, 0, 2, 0, 1, 0, 0, 0, 0, 0} \sim x^{-3\varepsilon}, \, x^0$   \\ 
 \hline
 $g_{54} \equiv s_{12} (s_{12} (1-x)-s_{23} x) \varepsilon^4 (1 + \varepsilon) F_{54}$ & $F_{54} \equiv G_{1, 0, 2, 1, 1, 1, 0, 1, 0, 1, 0, 0, 0, 0, 0} \sim x^0$   \\
 \hline
 $g_{55} \equiv s_{12} s_{23} x \varepsilon^5 F_{55}$ & $F_{55} \equiv G_{1, 0, 1, 2, 1, 1, 0, 1, 0, 1, 0, 0, 0, 0, 0} \sim x^0$   \\ 
 \hline
 $g_{56} \equiv (s_{12} + s_{23} x) \varepsilon^6 F_{56}$ & $F_{56} \equiv G_{1, 1, 0, 1, 1, 1, 0, 0, 1, 1, 0, 0, 0, 0, 0} \sim x^0$    \\
 \hline
 $g_{57} \equiv s_{12}^2 (1-x) \varepsilon^4 (1 + \varepsilon) F_{57}$ & $F_{57} \equiv G_{1, 1, 0, 2, 1, 1, 0, 0, 1, 1, 0, 0, 0, 0, 0} \sim x^{-3\varepsilon}, \, x^0$   \\ 
 \hline
 $g_{58} \equiv s_{23} x \varepsilon^6 F_{58}$ & $F_{58} \equiv G_{0, 1, 0, 1, 1, 1, 1, 0, 1, 1, 0, 0, 0, 0, 0} \sim x^{-3\varepsilon}, \, x^0$    \\
 \hline
 $g_{59} \equiv -s_{12}^2 x \varepsilon^5 F_{59}$ & $F_{59} \equiv G_{0, 2, 0, 1, 1, 1, 1, 0, 1, 1, 0, 0, 0, 0, 0} \sim x^{-3\varepsilon}, \, x^0$    \\
 \hline
 $g_{60} \equiv s_{12} (s_{12} (1-x) - s_{23} x) \varepsilon^5 F_{60}$ & $F_{60} \equiv G_{1, 0, 1, 1, 2, 1, 0, 0, 1, 1, 0, 0, 0, 0, 0} \sim x^{-3\varepsilon}, \, x^0$    \\ 
 \hline
 $g_{61} \equiv s_{12} (s_{12} (1-x)-s_{23} x) \varepsilon^5 F_{61}$ & $F_{61} \equiv G_{0, 1, 1, 2, 1, 1, 0, 0, 1, 1, 0, 0, 0, 0, 0} \sim x^{-3\varepsilon}, \, x^0$   \\
 \hline
 $g_{62} \equiv s_{12}^2 s_{23} x \varepsilon^5 F_{62}$ & $F_{62} \equiv G_{1, 1, 0, 1, 1, 2, 1, 1, 0, 1, 0, 0, 0, 0, 0} \sim x^{-1-3\varepsilon}, \, x^{-3\varepsilon}, \, x^0$    \\ 
 \hline
 $g_{63} \equiv s_{12}^2 x \varepsilon^5 (2 \varepsilon-1) F_{63}$ & $F_{63} \equiv G_{1, 1, 0, 1, 1, 1, 1, 1, 0, 1, 0, 0, 0, 0, 0} \sim x^{-3\varepsilon}, \, x^0$    \\
 \hline
 $g_{64} \equiv -s_{12} (s_{12} + s_{23}) x \varepsilon^6 F_{64}$ & $F_{64} \equiv G_{1, 0, 1, 1, 1, 1, 1, 1, 0, 1, 0, 0, 0, 0, 0} \sim x^{-3\varepsilon}, \, x^0$   \\ 
 \hline
 $g_{65} \equiv s_{12}^2 s_{23} x \varepsilon^5 F_{65}$ & $F_{65} \equiv G_{1, 0, 1, 1, 2, 1, 1, 1, 0, 1, 0, 0, 0, 0, 0} \sim x^{-1-3\varepsilon}, \, x^{-3\varepsilon}, \, x^0$   \\
 \hline
 $g_{66} \equiv -s_{12} (s_{12} + s_{23}) x \varepsilon^6 F_{66}$ & $F_{66} \equiv G_{1, 1, 0, 1, 1, 1, 1, 0, 1, 1, 0, 0, 0, 0, 0} \sim x^{-3\varepsilon}, \, x^0$   \\
 \hline
 $g_{67} \equiv s_{12}^2 s_{23} x \varepsilon^5 F_{67}$ & $F_{67} \equiv G_{1, 1, 0, 1, 2, 1, 1, 0, 1, 1, 0, 0, 0, 0, 0} \sim x^{-1-3\varepsilon}, \, x^{-3\varepsilon}, \, x^0$   \\ 
 \hline
 $g_{68} \equiv s_{12}^2 x \varepsilon^5 (2 \varepsilon-1) F_{68}$ & $F_{68} \equiv G_{1, 0, 1, 1, 1, 1, 1, 0, 1, 1, 0, 0, 0, 0, 0} \sim x^{-3\varepsilon}, \, x^0$  \\
 \hline
 $g_{69} \equiv s_{12}^2 s_{23} x \varepsilon^5 F_{69}$ & $F_{69} \equiv G_{1, 0, 1, 1, 2, 1, 1, 0, 1, 1, 0, 0, 0, 0, 0} \sim x^{-1-3\varepsilon}, \, x^{-3\varepsilon}, \, x^0$   \\ 
 \hline
 $g_{70} \equiv -s_{12} (s_{12} + s_{23}) x \varepsilon^6 F_{70}$ & $F_{70} \equiv G_{0, 1, 1, 1, 1, 1, 1, 0, 1, 1, 0, 0, 0, 0, 0} \sim x^{-3\varepsilon}, \, x^0$   \\
 \hline
 $g_{71} \equiv s_{12}^2 s_{23} x \varepsilon^5 F_{71}$ & $F_{71} \equiv G_{0, 1, 1, 2, 1, 1, 1, 0, 1, 1, 0, 0, 0, 0, 0} \sim x^{-1-3\varepsilon}, \, x^{-3\varepsilon}, \, x^0$   \\ 
 \hline
 $g_{72} \equiv s_{12} (s_{12} + s_{23} x) \varepsilon^6 F_{72}$ & $F_{72} \equiv G_{1, 0, 1, 1, 1, 1, 0, 1, 1, 1, 0, 0, 0, 0, 0} \sim x^{-3\varepsilon}, \, x^0$   \\
 \hline
 $g_{73} \equiv s_{12}^3 (1-x) \varepsilon^5 F_{73}$ & $F_{73} \equiv G_{2, 0, 1, 1, 1, 1, 0, 1, 1, 1, 0, 0, 0, 0, 0} \sim x^{-3\varepsilon}, \, x^0$   \\
 \hline
 $g_{74} \equiv s_{12}^2 s_{23} (1-x) x \varepsilon^5 F_{74}$ & $F_{74} \equiv G_{1, 0, 1, 1, 1, 1, 0, 1, 1, 2, 0, 0, 0, 0, 0} \sim x^{-1-3\varepsilon}, \, x^{-3\varepsilon}, \, x^0$   \\ 
 \hline
 $g_{75} \equiv s_{12}^2 s_{23} x \varepsilon^5 F_{75}$ & $F_{75} \equiv G_{1, 0, 1, 2, 1, 1, 0, 1, 1, 1, 0, 0, 0, 0, 0} \sim x^{-1-3\varepsilon}, \, x^{-3\varepsilon}, \, x^0$   \\
 \hline
 $g_{76} \equiv s_{12}^2 (1 - 2 \varepsilon) \varepsilon^5 F_{76}$ & $F_{76} \equiv G_{0, 1, 1, 1, 1, 1, 0, 1, 1, 1, 0, 0, 0, 0, 0} \sim x^{-3\varepsilon}, \, x^0$   \\ 
 \hline
 $g_{77} \equiv s_{12}^2 s_{23} x \varepsilon^5 F_{77}$ & $F_{77} \equiv G_{0, 1, 1, 2, 1, 1, 0, 1, 1, 1, 0, 0, 0, 0, 0} \sim x^{-1-3\varepsilon}, \, x^{-3\varepsilon}, \, x^0$   \\
 \hline
 $g_{78} \equiv s_{12}^2 (s_{12} (1-x)-s_{23} x) \varepsilon^6 F_{78}$ & $F_{78} \equiv G_{1, 1, 1, 1, 1, 1, 0, 1, 1, 1, 0, 0, 0, 0, 0} \sim x^{-3\varepsilon}, \, x^0$   \\ 
 \hline
$
g_{79} \equiv s_{12}^2 \varepsilon^5 (1 - 2 \varepsilon) F_{79} +g_{79}^{\text{rest}}$ & $F_{79} \equiv G_{1, 1, 1, 1, 1, 1, 0, 1, 1, 1, -1, 0, 0, 0, 0} \sim x^0$   \\
 \hline
 $g_{80} \equiv -s_{12}^2 x \varepsilon^6 F_{80}$ & $F_{80} \equiv G_{1, 1, 1, 1, 1, 1, 0, 1, 1, 1, 0, -1, 0, 0, 0} \sim x^0$   \\
 \hline
 $g_{81} \equiv s_{12}^3 s_{23} x \varepsilon^6 F_{81}$ & $F_{81} \equiv G_{1, 1, 1, 1, 1, 1, 1, 1, 1, 1, 0, 0, 0, 0, 0} \sim x^{-1-3\varepsilon}, \, x^{-3\varepsilon}, \, x^0$   \\ 
 \hline
 $g_{82} \equiv -s_{12}^3 x \varepsilon^6 F_{82}$ & $F_{82} \equiv G_{1, 1, 1, 1, 1, 1, 1, 1, 1, 1, -1, 0, 0, 0, 0} \sim x^{-3\varepsilon}, \, x^0$   \\
 \hline
 $g_{83} \equiv -s_{12}^3 x \varepsilon^6 F_{83}$ & $F_{83} \equiv G_{1, 1, 1, 1, 1, 1, 1, 1, 1, 1, 0, -1, 0, 0, 0} \sim x^{-3\varepsilon}, \, x^0$  \\ 
 \hline
\end{tabular}
\end{center}
\end{adjustwidth}
\caption{The rest 41 U.T. basis elements in the SDE notation. For space convenience we used the notation $g_{79}^{\text{rest}} \equiv s_{12} (x-1) \varepsilon^5 (2 F_{23} + 3 F_{25} - 4 F_{26} + F_{28}-2 s_{12} F_{45} + 2 \varepsilon F_{47} + 2 s_{12} \varepsilon F_{80}) -  2 x s_{12} s_{23} \varepsilon^6 F_{72}$.}
\label{table:2}
\end{table}


\begin{thebibliography}{110}


\bibitem{de1}
A.~V.~Kotikov,
``Differential equations method: New technique for massive Feynman diagrams calculation,''
\href{https://www.sciencedirect.com/science/article/abs/pii/037026939190413K?via}{Phys. Lett. B \textbf{254} (1991), 158-164}.

\bibitem{de2}
A.~V.~Kotikov,
``Differential equations method: The Calculation of vertex type Feynman diagrams,''
\href{https://www.sciencedirect.com/science/article/abs/pii/037026939190834D?via}{Phys. Lett. B \textbf{259} (1991), 314-322}.

\bibitem{de3}
A.~V.~Kotikov,
``Differential equation method: The Calculation of N point Feynman diagrams,''
\href{https://www.sciencedirect.com/science/article/abs/pii/037026939190536Y?via}{Phys. Lett. B \textbf{267} (1991), 123-127} [Errattum: \href{https://www.sciencedirect.com/science/article/pii/037026939291582T?via}{Phys. Lett. B \textbf{295} (1992), 409}].

\bibitem{de4}
T.~Gehrmann and E.~Remiddi,
``Differential equations for two loop four point functions,''
Nucl. Phys. B \textbf{580} (2000), 485-518
\href{https://arxiv.org/abs/hep-ph/9912329}{[arXiv:hep-ph/9912329 [hep-ph]]}.


\bibitem{IBP1}
K.~G.~Chetyrkin and F.~V.~Tkachov,
``Integration by Parts: The Algorithm to Calculate beta Functions in 4 Loops,''
\href{https://www.sciencedirect.com/science/article/abs/pii/0550321381901991?via}{Nucl. Phys. B \textbf{192} (1981), 159-204}.

\bibitem{kira}
J.~Klappert, F.~Lange, P.~Maierhofer and J.~Usovitsch,
``Integral Reduction with Kira 2.0 and Finite Field Methods,''
\href{https://arxiv.org/abs/2008.06494}{[arXiv:2008.06494 [hep-ph]]}.

\bibitem{fire}
A.~V.~Smirnov and F.~S.~Chuharev,
``FIRE6: Feynman Integral REduction with Modular Arithmetic,''
\href{https://arxiv.org/abs/1901.07808}{[arXiv:1901.07808 [hep-ph]]}.

\bibitem{litered}
R.~N.~Lee,
``LiteRed 1.4: a powerful tool for reduction of multiloop integrals,''
J. Phys. Conf. Ser. \textbf{523} (2014), 012059
\href{https://arxiv.org/pdf/1310.1145.pdf}{[arXiv:1310.1145 [hep-ph]]}.

\bibitem{IBP2}
S.~Laporta,
``High precision calculation of multiloop Feynman integrals by difference equations,''
Int. J. Mod. Phys. A \textbf{15} (2000), 5087-5159
\href{https://arxiv.org/abs/hep-ph/0102033}{[arXiv:hep-ph/0102033 [hep-ph]]}.

\bibitem{IBP3}
K.~J.~Larsen and Y.~Zhang,
``Integration-by-parts reductions from unitarity cuts and algebraic geometry,''
Phys. Rev. D \textbf{93} (2016) no.4, 041701
\href{https://arxiv.org/abs/1511.01071}{[arXiv:1511.01071 [hep-th]]}.

\bibitem{IBP4}
D.~Bendle, J.~Bohm, W.~Decker, A.~Georgoudis, F.~J.~Pfreundt, M.~Rahn, P.~Wasser and Y.~Zhang,
``Integration-by-parts reductions of Feynman integrals using Singular and GPI-Space,''
JHEP \textbf{02} (2020), 079
\href{https://arxiv.org/abs/1908.04301}{[arXiv:1908.04301 [hep-th]]}.

\bibitem{IBP5}
D.~A.~Kosower,
``Direct Solution of Integration-by-Parts Systems,''
Phys. Rev. D \textbf{98} (2018) no.2, 025008
doi:10.1103/PhysRevD.98.025008
\href{https://arxiv.org/abs/1804.00131}{[arXiv:1804.00131 [hep-ph]]}.

\bibitem{IBP6}
P.~Mastrolia and S.~Mizera,
``Feynman Integrals and Intersection Theory,''
JHEP \textbf{02} (2019), 139
\href{https://arxiv.org/abs/1810.03818}{[arXiv:1810.03818 [hep-th]]}.

\bibitem{IBP7}
H.~Frellesvig, F.~Gasparotto, S.~Laporta, M.~K.~Mandal, P.~Mastrolia, L.~Mattiazzi and S.~Mizera,
``Decomposition of Feynman Integrals on the Maximal Cut by Intersection Numbers,''
JHEP \textbf{05} (2019), 153
\href{https://arxiv.org/abs/1901.11510}{[arXiv:1901.11510 [hep-ph]]}.

\bibitem{IBP8}
H.~Frellesvig, F.~Gasparotto, M.~K.~Mandal, P.~Mastrolia, L.~Mattiazzi and S.~Mizera,
``Vector Space of Feynman Integrals and Multivariate Intersection Numbers,''
Phys. Rev. Lett. \textbf{123} (2019) no.20, 201602
\href{https://arxiv.org/abs/1907.02000}{[arXiv:1907.02000 [hep-th]]}.

\bibitem{IBP9}
H.~Frellesvig, F.~Gasparotto, S.~Laporta, M.~K.~Mandal, P.~Mastrolia, L.~Mattiazzi and S.~Mizera,
``Decomposition of Feynman Integrals by Multivariate Intersection Numbers,''
\href{https://arxiv.org/abs/2008.04823}{[arXiv:2008.04823 [hep-th]]}.

\bibitem{IBP10}
J.~Klappert and F.~Lange,
``Reconstructing rational functions with FireFly,''
Comput. Phys. Commun. \textbf{247} (2020), 106951
\href{https://arxiv.org/abs/1904.00009}{[arXiv:1904.00009 [cs.SC]]}.

\bibitem{IBP11}
J.~Klappert, S.~Y.~Klein and F.~Lange,
``Interpolation of Dense and Sparse Rational Functions and other Improvements in $\texttt{FireFly}$,''
\href{https://arxiv.org/abs/2004.01463}{[arXiv:2004.01463 [cs.MS]]}.


\bibitem{Henn1}
J.~M.~Henn,
``Multiloop integrals in dimensional regularization made simple,''
Phys. Rev. Lett. \textbf{110} (2013), 251601
\href{https://arxiv.org/abs/1304.1806}{[arXiv:1304.1806 [hep-th]]}.

\bibitem{Henn2}
J.~M.~Henn,
``Lectures on differential equations for Feynman integrals,''
J. Phys. A \textbf{48} (2015), 153001
\href{https://arxiv.org/abs/1412.2296}{[arXiv:1412.2296 [hep-ph]]}.

\bibitem{Henn2}
J.~M.~Henn, K.~Melnikov and V.~A.~Smirnov,
``Two-loop planar master integrals for the production of off-shell vector bosons in hadron collisions,''
JHEP \textbf{05} (2014), 090
doi:10.1007/JHEP05(2014)090
\href{https://arxiv.org/abs/1402.7078}{[arXiv:1402.7078 [hep-ph]]}.

\bibitem{CB1}
R.~N.~Lee,
``Reducing differential equations for multiloop master integrals,''
JHEP \textbf{04} (2015), 108
\href{https://arxiv.org/abs/1411.0911}{[arXiv:1411.0911 [hep-ph]]}.

\bibitem{CB2}
M.~Prausa,
``epsilon: A tool to find a canonical basis of master integrals,''
Comput. Phys. Commun. \textbf{219} (2017), 361-376
\href{https://arxiv.org/abs/1701.00725}{[arXiv:1701.00725 [hep-ph]]}.

\bibitem{CB3}
O.~Gituliar and V.~Magerya,
``Fuchsia: a tool for reducing differential equations for Feynman master integrals to epsilon form,''
Comput. Phys. Commun. \textbf{219} (2017), 329-338
\href{https://arxiv.org/abs/1701.04269}{[arXiv:1701.04269 [hep-ph]]}.

\bibitem{CB4}
C.~Meyer,
``Algorithmic transformation of multi-loop master integrals to a canonical basis with CANONICA,''
Comput. Phys. Commun. \textbf{222} (2018), 295-312
\href{https://arxiv.org/abs/1705.06252}{[arXiv:1705.06252 [hep-ph]]}.

\bibitem{CB5}
S.~Abreu, B.~Page and M.~Zeng,
``Differential equations from unitarity cuts: nonplanar hexa-box integrals,''
JHEP \textbf{01} (2019), 006
\href{https://arxiv.org/abs/1807.11522}{[arXiv:1807.11522 [hep-th]]}.

\bibitem{CB6}
D.~Chicherin, T.~Gehrmann, J.~M.~Henn, P.~Wasser, Y.~Zhang and S.~Zoia,
``All Master Integrals for Three-Jet Production at Next-to-Next-to-Leading Order,''
Phys. Rev. Lett. \textbf{123} (2019) no.4, 041603
\href{https://arxiv.org/abs/1812.11160}{[arXiv:1812.11160 [hep-ph]]}.

\bibitem{CB7}
C.~Dlapa, J.~Henn and K.~Yan,
``Deriving canonical differential equations for Feynman integrals from a single uniform weight integral,''
JHEP \textbf{05} (2020), 025
\href{https://arxiv.org/abs/2002.02340}{[arXiv:2002.02340 [hep-ph]]}.

\bibitem{CB8}
J.~Henn, B.~Mistlberger, V.~A.~Smirnov and P.~Wasser,
``Constructing d-log integrands and computing master integrals for three-loop four-particle scattering,''
JHEP \textbf{04} (2020), 167
\href{https://arxiv.org/abs/2002.09492}{[arXiv:2002.09492 [hep-ph]]}.

\bibitem{CB9}
S.~Abreu, H.~Ita, F.~Moriello, B.~Page, W.~Tschernow and M.~Zeng,
``Two-Loop Integrals for Planar Five-Point One-Mass Processes,''
\href{https://arxiv.org/abs/2005.04195}{[arXiv:2005.04195 [hep-ph]]}.


\bibitem{SDE1}
C.~G.~Papadopoulos,
``Simplified differential equations approach for Master Integrals,''
JHEP \textbf{07} (2014), 088
\href{https://arxiv.org/abs/1401.6057}{[arXiv:1401.6057 [hep-ph]]}.

\bibitem{SDE2}
C.~G.~Papadopoulos, D.~Tommasini and C.~Wever,
``Two-loop Master Integrals with the Simplified Differential Equations approach,''
JHEP \textbf{01} (2015), 072
\href{https://arxiv.org/abs/1409.6114}{[arXiv:1409.6114 [hep-ph]]}.

\bibitem{SDE3}
C.~G.~Papadopoulos, D.~Tommasini and C.~Wever,
``The Pentabox Master Integrals with the Simplified Differential Equations approach,''
JHEP \textbf{04} (2016), 078
\href{https://arxiv.org/abs/1511.09404}{[arXiv:1511.09404 [hep-ph]]}.


\bibitem{GPL}
A.~B.~Goncharov,
``Multiple polylogarithms, cyclotomy and modular complexes,''
Math. Res. Lett. \textbf{5} (1998), 497-516
\href{https://arxiv.org/abs/1105.2076}{[arXiv:1105.2076 [math.AG]]}.


\bibitem{mastrolia}
S.~Di Vita, P.~Mastrolia, U.~Schubert and V.~Yundin,
``Three-loop master integrals for ladder-box diagrams with one massive leg,''
JHEP \textbf{09} (2014), 148
\href{https://arxiv.org/abs/1408.3107}{[arXiv:1408.3107 [hep-ph]]}.

\bibitem{Henn:2013tua}
J.~M.~Henn, A.~V.~Smirnov and V.~A.~Smirnov,
``Analytic results for planar three-loop four-point integrals from a Knizhnik-Zamolodchikov equation,''
JHEP \textbf{07} (2013), 128
doi:10.1007/JHEP07(2013)128
\href{https://arxiv.org/abs/1306.2799}{[arXiv:1306.2799 [hep-th]]}.


\bibitem{jaxodraw}
D.~Binosi, J.~Collins, C.~Kaufhold and L.~Theussl,
``JaxoDraw: A Graphical user interface for drawing Feynman diagrams. Version 2.0 release notes,''
Comput. Phys. Commun. \textbf{180} (2009), 1709-1715
\href{https://arxiv.org/abs/0811.4113}{[arXiv:0811.4113 [hep-ph]]}.



\bibitem{Polylogtools}
C.~Duhr and F.~Dulat,
``PolyLogTools - polylogs for the masses,''
JHEP \textbf{08} (2019), 135
\href{https://arxiv.org/abs/1904.07279}{[arXiv:1904.07279 [hep-th]]}.

\bibitem{hyper}
E.~Panzer,
``Algorithms for the symbolic integration of hyperlogarithms with applications to Feynman integrals,''
Comput. Phys. Commun. \textbf{188} (2015), 148-166
\href{https://arxiv.org/abs/1403.3385}{[arXiv:1403.3385 [hep-th]]}.


\bibitem{asy1}
M.~Beneke and V.~A.~Smirnov,
``Asymptotic expansion of Feynman integrals near threshold,''
Nucl. Phys. B \textbf{522} (1998), 321-344
\href{https://arxiv.org/abs/hep-ph/9711391}{[arXiv:hep-ph/9711391 [hep-ph]]}.

\bibitem{asy2}
V.~A.~Smirnov,
``Problems of the strategy of regions,''
Phys. Lett. B \textbf{465} (1999), 226-234
\href{https://arxiv.org/abs/hep-ph/9907471}{[arXiv:hep-ph/9907471 [hep-ph]]}.

\bibitem{asy3}
A.~Pak and A.~Smirnov,
``Geometric approach to asymptotic expansion of Feynman integrals,''
Eur. Phys. J. C \textbf{71} (2011), 1626
\href{https://arxiv.org/abs/1011.4863}{[arXiv:1011.4863 [hep-ph]]}.

\bibitem{asy4}
B.~Jantzen,
``Foundation and generalization of the expansion by regions,''
JHEP \textbf{12} (2011), 076
\href{https://arxiv.org/abs/1111.2589}{[arXiv:1111.2589 [hep-ph]]}.

\bibitem{asy5}
B.~Jantzen, A.~V.~Smirnov and V.~A.~Smirnov,
``Expansion by regions: revealing potential and Glauber regions automatically,''
Eur. Phys. J. C \textbf{72} (2012), 2139
\href{https://arxiv.org/abs/1206.0546}{[arXiv:1206.0546 [hep-ph]]}.

\bibitem{asy6}
T.~Y.~Semenova, A.~V.~Smirnov and V.~A.~Smirnov,
``On the status of expansion by regions,''
Eur. Phys. J. C \textbf{79} (2019) no.2, 136
\href{https://arxiv.org/abs/1809.04325}{[arXiv:1809.04325 [hep-th]]}.


\bibitem{fiesta}
A.~V.~Smirnov,
``FIESTA4: Optimized Feynman integral calculations with GPU support,''
Comput. Phys. Commun. \textbf{204} (2016), 189-199
\href{https://arxiv.org/abs/1511.03614}{[arXiv:1511.03614 [hep-ph]]}.

\bibitem{pysecdec}
S.~Borowka, G.~Heinrich, S.~Jahn, S.~P.~Jones, M.~Kerner, J.~Schlenk and T.~Zirke,
``pySecDec: a toolbox for the numerical evaluation of multi-scale integrals,''
Comput. Phys. Commun. \textbf{222} (2018), 313-326
\href{https://arxiv.org/abs/1703.09692}{[arXiv:1703.09692 [hep-ph]]}.
 
 
\bibitem{Syrrakosthesis}
N.~Syrrakos, PhD thesis, \textit{in preparation}.
 
\bibitem{5box}
D.~D.~Canko, C.~G.~Papadopoulos and N.~Syrrakos,
``Analytic representation of all planar two-loop five-point Master Integrals with one off-shell leg,''
\href{https://arxiv.org/abs/2009.13917}{[arXiv:2009.13917 [hep-ph]]}.

\bibitem{Papadopoulos:2019iam}
C.~G.~Papadopoulos and C.~Wever,
``Internal Reduction method for computing Feynman Integrals,''
JHEP \textbf{02} (2020), 112
doi:10.1007/JHEP02(2020)112
\href{https://arxiv.org/abs/1910.06275}{[arXiv:1910.06275 [hep-ph]]}.
 
 
\bibitem{Duhr}
C.~Duhr,
``Mathematical aspects of scattering amplitudes,''
\href{https://arxiv.org/abs/1411.7538}{[arXiv:1411.7538 [hep-ph]]}. 

\bibitem{Moriello1}
F.~Moriello,
``Generalised power series expansions for the elliptic planar families of Higgs + jet production at two loops,''
JHEP \textbf{01} (2020), 150
\href{https://arxiv.org/abs/1907.13234}{[arXiv:1907.13234 [hep-ph]]}.

\bibitem{Moriello2}
R.~Bonciani, V.~Del Duca, H.~Frellesvig, J.~M.~Henn, M.~Hidding, L.~Maestri, F.~Moriello, G.~Salvatori and V.~A.~Smirnov,
``Evaluating a family of two-loop non-planar master integrals for Higgs + jet production with full heavy-quark mass dependence,''
JHEP \textbf{01} (2020), 132
\href{https://arxiv.org/abs/1907.13156}{[arXiv:1907.13156 [hep-ph]]}.

\bibitem{Moriello3}
H.~Frellesvig, M.~Hidding, L.~Maestri, F.~Moriello and G.~Salvatori,
``The complete set of two-loop master integrals for Higgs + jet production in QCD,''
JHEP \textbf{06} (2020), 093
\href{https://arxiv.org/abs/1911.06308}{[arXiv:1911.06308 [hep-ph]]}.

\bibitem{Hidding}
M.~Hidding,
``DiffExp, a Mathematica package for computing Feynman integrals in terms of one-dimensional series expansions,''
\href{https://arxiv.org/abs/2006.05510}{[arXiv:2006.05510 [hep-ph]]}.

\bibitem{Anastasiou1}
C.~Anastasiou and G.~Sterman,
``Removing infrared divergences from two-loop integrals,''
JHEP \textbf{07}, 056 (2019)
\href{https://arxiv.org/abs/1812.03753}{[arXiv:1812.03753 [hep-ph]]}.

\bibitem{Anastasiou2}
C.~Anastasiou, R.~Haindl, G.~Sterman, Z.~Yang and M.~Zeng,
``Locally finite two-loop amplitudes for off-shell multi-photon production in electron-positron annihilation,''
\href{https://arxiv.org/abs/2008.12293}{[arXiv:2008.12293 [hep-ph]]}.

\bibitem{Heinrich}
G.~Heinrich,
``Collider Physics at the Precision Frontier,''
\href{https://arxiv.org/abs/2009.00516}{[arXiv:2009.00516 [hep-ph]]}.

\bibitem{Amoroso:2020lgh}
S.~Amoroso, P.~Azzurri, J.~Bendavid, E.~Bothmann, D.~Britzger, H.~Brooks, A.~Buckley, M.~Calvetti, X.~Chen and M.~Chiesa, \textit{et al.}
``Les Houches 2019: Physics at TeV Colliders: Standard Model Working Group Report,''
\href{https://arxiv.org/abs/2003.01700}{[arXiv:2003.01700 [hep-ph]]}.


\end{thebibliography}
\end{document}